\documentclass[twoside,twocolumn,9pt]{article}
\usepackage{extsizes}
\usepackage[super,sort&compress,comma]{natbib} 
\usepackage[version=3]{mhchem}
\usepackage[left=1.5cm, right=1.5cm, top=1.785cm, bottom=2.0cm]{geometry}
\usepackage{balance}
\usepackage{times,mathptmx}
\usepackage{sectsty}
\usepackage{graphicx} 
\usepackage{lastpage}
\usepackage{dblfloatfix}
\usepackage[format=plain,justification=justified,singlelinecheck=false,font={stretch=1.125,small,sf},labelfont=bf,labelsep=space]{caption}
\usepackage{float}
\usepackage{fancyhdr}
\usepackage{fnpos}
\usepackage[english]{babel}
\usepackage{array}
\usepackage{droidsans}
\usepackage{charter}
\usepackage[T1]{fontenc}
\usepackage[usenames,dvipsnames]{xcolor}
\usepackage{setspace}
\usepackage[compact]{titlesec}

\usepackage{epstopdf}

\definecolor{cream}{RGB}{222,217,201}
\usepackage{hyperref}
\hypersetup{backref,colorlinks=true,linkcolor=blue,citecolor=blue,urlcolor=blue,linktocpage=true,breaklinks}
\begin{document}

\pagestyle{fancy}
\thispagestyle{plain}
\fancypagestyle{plain}{

\renewcommand{\headrulewidth}{0pt}
}

\makeFNbottom
\makeatletter
\renewcommand\LARGE{\@setfontsize\LARGE{15pt}{17}}
\renewcommand\Large{\@setfontsize\Large{12pt}{14}}
\renewcommand\large{\@setfontsize\large{10pt}{12}}
\renewcommand\footnotesize{\@setfontsize\footnotesize{7pt}{10}}
\makeatother

\renewcommand{\thefootnote}{\fnsymbol{footnote}}
\renewcommand\footnoterule{\vspace*{1pt}%
\color{cream}\hrule width 3.5in height 0.4pt \color{black}\vspace*{5pt}} 
\setcounter{secnumdepth}{5}

\makeatletter 
\renewcommand\@biblabel[1]{#1}            
\renewcommand\@makefntext[1]%
{\noindent\makebox[0pt][r]{\@thefnmark\,}#1}
\makeatother 
\renewcommand{\figurename}{\small{Fig.}~}
\sectionfont{\sffamily\large}
\subsectionfont{\normalsize}
\subsubsectionfont{\bf}
\setstretch{1.125} 
\setlength{\skip\footins}{0.8cm}
\setlength{\footnotesep}{0.25cm}
\setlength{\jot}{10pt}
\titlespacing*{\section}{0pt}{4pt}{4pt}
\titlespacing*{\subsection}{0pt}{15pt}{1pt}

\fancyfoot{}
\fancyfoot[RO]{\footnotesize{\sffamily{1--\pageref{LastPage} ~\textbar  \hspace{2pt}\thepage}}}
\fancyfoot[LE]{\footnotesize{\sffamily{\thepage~\textbar\hspace{4.65cm} 1--\pageref{LastPage}}}}
\fancyhead{}
\renewcommand{\headrulewidth}{0pt} 
\renewcommand{\footrulewidth}{0pt}
\setlength{\arrayrulewidth}{1pt}
\setlength{\columnsep}{6.5mm}
\setlength\bibsep{1pt}

\makeatletter 
\newlength{\figrulesep} 
\setlength{\figrulesep}{0.5\textfloatsep} 

\newcommand{\topfigrule}{\vspace*{-1pt}%
\noindent{\color{cream}\rule[-\figrulesep]{\columnwidth}{1.5pt}} }

\newcommand{\botfigrule}{\vspace*{-2pt}%
\noindent{\color{cream}\rule[\figrulesep]{\columnwidth}{1.5pt}} }

\newcommand{\dblfigrule}{\vspace*{-1pt}%
\noindent{\color{cream}\rule[-\figrulesep]{\textwidth}{1.5pt}} }

\makeatother

\twocolumn[
  \begin{@twocolumnfalse}
\vspace{0.5cm}
\sffamily
\begin{tabular}{m{0.5cm} p{16.5cm} }

 & \noindent\LARGE{\textbf{Bound and continuum-embedded states of cyano\-polyyne anions$^\dag$}} \\
\vspace{0.2cm} & \vspace{0.1cm} \\

 & \noindent\large{Wojciech Skomorowski, Sahil Gulania, and Anna I. Krylov\textit{$^{a}$}\vspace{0.3cm}} \\

 & \noindent\normalsize{Cyanopolyyne anions were among the first anions discovered in the                
  interstellar medium. The discovery have raised questions                         
  about routes of formation of these anions in space.                              
  Some of the proposed mechanisms assumed that anionic excited electronic          
  states, either metastable or weakly bound, play a key role in the formation  process.
  Verification of this hypothesis requires  detailed knowledge of the electronic states of
  the anions.                                                                      
  Here we investigate bound and continuum states of four  cyanopolyyne             
  anions, CN$^-$, C$_3$N$^-$, C$_5$N$^-$, and  C$_7$N$^-$,                         
  by means of  {\it ab initio}  calculations. We  employ the                       
  equation-of-motion coupled-cluster method augmented with complex absorbing       
  potential.                                                                       
  We predict that already in CN$^-$, the smallest anion in the family,             
  there are several low-lying metastable states of both singlet and triplet spin symmetry.
  These states, identified as shape resonances,                                    
  are located between 6.3--8.5 eV above the                                         
  ground state of the anion                                                        
  (or 2.3--4.5 eV above the ground state of the parent radical) and                 
  have widths of a few tenths of eV up to 1 eV. We analyze the identified resonances
  in terms of leading molecular orbital contributions and  Dyson orbitals.         
  As the carbon chain length increases in the C$_{2n+1}$N$^-$ series, these resonances
  gradually become stabilized and eventually turn into stable valence bound states.
  The trends in energies of the transitions leading to both resonance              
  and bound excited states can be rationalized by means of the  H\"{u}ckel         
  model. Apart from valence excited states, some of the cyanopolyynes can also support     
  dipole bound states and dipole stabilized resonances, owing to a                 
  large dipole moment of the parent radicals in the lowest $^2\Sigma^+$ state.     
  We discuss the consequences of  open-shell character of the neutral radicals     
  on the dipole-stabilized states of the respective anions.} \\

\end{tabular}

 \end{@twocolumnfalse} \vspace{0.6cm}

  ]

\renewcommand*\rmdefault{bch}\normalfont\upshape
\rmfamily
\section*{}
\vspace{-1cm}


\footnotetext{\textit{$^{a}$~Department of Chemistry, University of Southern California, Los Angeles, California 90089, USA; E-mail: krylov@usc.edu (A.I.K.), skomorow@usc.edu (W.S.).}}

\footnotetext{\dag~Electronic Supplementary Information (ESI) available: CAP parameters, $\eta$ trajectories, details of complex basis function and cross section calculations,
  relevant Cartesian geometries and basis set details.}

\section{Introduction}
Electronic spectra of typical  anions are markedly different from those of       
  neutral molecules and cations\cite{Simons:08:MolAnions,Simons:ARPC:11,Simons:book:00,Herbert:RCC:15}.
  Neutral and positively charged species have  numerous electronically excited bound
  states, particularly  near the ionization threshold where an infinite series of  
  Rydberg states appears. Yet typical anions only support a few (if any) electronically bound     
  states\cite{Garrett:JCP:82,Gutsev:CPL:97,Sommerfeld:PCCP:02}.                    
  Electronic states of anions can be broadly classified into the following         
  categories: valence,  dipole-bound,                                              
  and correlation-bound states\cite{Simons:08:MolAnions,Simons:book:00,Herbert:RCC:15,Sommerfeld:NaCl:10,Voora:14,Voora:C60:2013}.
  Bound excited valence states are rare  due to relatively low  electron detachment energy.
  Dipole-bound states (DBS) are reminiscent of Rydberg states in neutral systems,  
  however, due to a different type of long-range interaction (charge -- dipole     
  instead of long--range Coulomb potential), DBS are less ubiquitous and  only     
  appear when the neutral precursor possesses sufficiently large dipole            
  moment\cite{Simons:08:MolAnions,Herbert:RCC:15,Jordan:DBRev:03,Gutovski:DBA:98,Crawford:DBS:11,Lykke:PRL:84}.
  A separate class of electronically excited states of anions comprises  auto-detaching
  resonances\cite{Herbert:RCC:15,Simons:RevAn:14}, i.e., states with finite lifetime located
  above the lowest electron detachment threshold;                                  
  those include singly excited states                                              
  (which can be shape or Feshbach type) and doubly excited                         
  Feshbach resonances derived by electron attachment to an electronically excited  
  neutral core\cite{FR}.                                                           
  Often such metastable states are the only (semi)-discrete features of anionic  spectra;
  thus, they  play an important role in their spectroscopic characterization\cite{Jordan:Acetylene:78,Dressler:Acetylene:87,Weinkauf:benzoQ:99}.
                                                                                   
  The discovery of simple anions in the interstellar medium (ISM) in the last      
  decade has stimulated interest in chemical properties of these                   
  species.                                                                         
  C$_6$H$^-$, identified in 2006\cite{McCarthy:AJL:06}, became the first known ISM anion.
  So far, five other molecular anions have been confirmed to be present in the ISM: C$_4$H$^-$, C$_8$H$^-$,
  CN$^-$, C$_3$N$^-$, and  C$_5$N$^-$\cite{Bruken:AJL:07,Thaddeus:AJ:08,Cernicharo:AA:07,Cernicharo:AJL:08,Agundez:AA:10}.
  These observations, possible routes of the formation of these anions, and        
  theoretical modeling of their abundances                                         
  are summarized in a recent review\cite{Millar:ChemRev:17}.                       
  Astrophysical discoveries of anions in the ISM were accompanied  by              
  laboratory measurements of rotational transitions                                
  and  quantum-chemical calculations, which were instrumental for the              
  correct assignment of the observed                                               
  spectra\cite{Gupta:AJL:07,Botschwina:JCP:08,McCarthy:JCP:08,Gottlieb:JCP:07,Aoki:CPL:00}.
  All interstellar  anions identified so far are closely related and belong to the class of
  linear polyynes (C$_{2n}$H$^-$) or                                               
  cyanopolyynes (C$_{2n-1}$N$^-$), species with the conjugated chains of           
  triple CC and CN bonds.                                                          
  Their ground states have closed-shell electronic                                 
  configuration ($^1\Sigma^+$), giving rise to large electron                      
  detachment  energy of about $\sim$4 eV. Several                                  
  theoretical studies of  (cyano)polyynes anions                                   
  focused on their ground-state spectroscopic                                      
  properties\cite{Wang:CPL:95,Zhan:JCP:96,Coupeaud:JCP:08,Botschwina:JCP:08,Stein:MP:15}.
  The most comprehensive study by Botschwina and Oswald\cite{Botschwina:JCP:08} reported
  equilibrium structures, dipole moments, vertical detachment                      
  energies, and  rotational--vibrational parameters for the C$_{2n-1}$N$^-$ anions 
  using the                                                                        
  highly accurate  CCSD(T) method (coupled-cluster with single, double and non-iterative
  triple excitations).                                                             
  In contrast to the well-studied                                                  
  rovibrational structure of the C$_{2n-1}$N$^-$ anions\cite{McCarthy:JCP:08,Gottlieb:JCP:07,Wang:CPL:95,Zhan:JCP:96,Coupeaud:JCP:08,Botschwina:JCP:08,Stein:MP:15,Kolos:JCP:08},
  their electronic spectra, in particular in the continuum part, have not been systematically
  investigated.                                                                    
  Better understanding  of excited states in these simple  anions is important not 
  only for the  astrochemical research, but also                                   
  in the context of precision spectroscopy\cite{Ito:PRL:14,Zhu:PRL:17} and recently proposed
  experiments with cold molecular anions\cite{Yzombard:PRL:15,Tomza:PCCP:17} to study
  properties of matter at the quantum limit,                                       
  where electronically excited states play a key role.
  
  Here we  characterize                                                            
  electronically excited bound and near-threshold continuum states of four cyanopolyyne anions
  C$_{2n-1}$N$^-\; (n=1,\dots, 4)$                                                 
  with high-level {\it ab initio} methods. The results are   particularly  relevant in the
  context of                                                                       
  the ongoing discussion on possible mechanisms of                                 
  their formation in the ISM\cite{Herbst:AJ:08,Carelli:AJ:13,Carelli:DBS:14,Satta:AJ:15,Douguet:CxNm:16}. One
  hypothesis is that the cyanopolyyne anions might be formed via direct            
  electron attachment to the respective neutral precursors (cyanopolyyne radicals) that are
  ubiquitous in the ISM, followed by spontaneous radiative  transition to the      
  anionic ground state.                                                            
  This radiative electron attachment (REA) process                                 
  has been studied recently in the three smallest                                  
  cyanopolyyne anions\cite{Douguet:CxNm:16}; the calculations showed that the direct REA rates
  appear too small to explain the abundance of these anions in the ISM.            
  The efficiency  of the electron capture process can be significantly enhanced by 
  metastable and weakly bound electronic states                                    
  in the vicinity of the electron detachment threshold. These  electronic states   
  can resonantly trap incident electrons for a finite time,  such that the         
  complex can be  stabilized and subsequently relax to the ground state            
  of the anion.                                                                    
  This mechanism can be described as                                               
  indirect radiative electron attachment\cite{Douguet:JCP:15}. Thus,  electronic or
  mixed electronic-vibrational resonances can act as a doorway towards efficient   
  formation of valence bound anions by providing a mechanism for initial           
  binding of the excess electron. A similar mechanism,                             
  electron attachment facilitated by the $p-$wave shape                            
  resonances, has been reported for                                                
  closed-shell acetylenic carbon-rich neutral molecules                            
  (such as NC$_4$N and HC$_3$N)                                                    
  \cite{Graupner:AJL:95,Sebastianelli:EPDJ:12,Sebastianelli:EPJD:10}.              
  The prospects for the                                                            
  enhancement of  REA via near-threshold excited anionic electronic states         
  were discussed in Ref. \citenum{Carelli:DBS:14}, where it was suggested       
  that REA may be particularly effective in strongly polar species that are likely to
  form near-threshold bound or metastable states.
  
  Several theoretical studies  attempted to investigate the  existence of  bound and
  metastable states in the shortest                                                
  C$_{2n-1}$N$^-$ anions. Harrison and Tennyson                                    
  performed electron scattering calculations for CN$+e^-$ and  C$_3$N$+e^-$ by using
  the $R-$matrix method\cite{Harrison:JPB:11,Harrison:JPB:12}; this study          
  reported several triplet resonances                                              
  in CN$^-$ (in the energy range from 3 to 5 eV above the electron detachment      
  threshold) and singlet and triplet resonances in C$_3$N$^-$ (from 0.7            
  to 3 eV above the threshold). Additionally, several very weakly bound $\Sigma^+$ 
  and $\Pi$ states have been claimed to exist in C$_3$N$^-$.                       
  No analysis in terms of orbital character or electronic configurations  has been given for
  aforementioned  resonances and  bound states\cite{Harrison:JPB:11,Harrison:JPB:12}.
  The report\cite{Harrison:JPB:11,Harrison:JPB:12} of                              
  numerous near-threshold bound states in C$_3$N$^-$ contradicts                   
  the findings of previous                                                         
  quantum-chemical calculations\cite{Crawford:DBS:11} performed with EOM-CCSD method
  (equation-of-motion CCSD)\cite{Krylov:EOMRev:07,Christiansen:EOMRev:11,Bartlet:EOMREV:12} in which
  no bound states were found in C$_3$N$^-$.                                        
  A later study\cite{Fortenberry:JPCA:15} revisited                                
  these EOM-CCSD  results and concluded that C$_3$N$^-$ and C$_5$N$^-$             
  support a single dipole-bound state.                                             
  No similar studies have been reported for longer C$_{2n-1}$N$^-$ chains.         
                                                                                   
  The most relevant experimental study\cite{Wester:interstellarAnDestruct:13}      
  of C$_{2n-1}$N$^-$                                                               
  is the measurement of photodetachment cross section                              
  for CN$^-$  and C$_3$N$^-$  at the wavelength of 266 nm (4.66 eV). The           
  authors concluded  that photodetachment by interstellar UV photons is the major  
  destruction  mechanism for these anions\cite{Wester:interstellarAnDestruct:13}.  
  Earlier  studies based on photoelectron spectra provided  accurate electron detachment
  energies for CN$^-$\cite{Bradforth:JCP:98},                                      
  C$_3$N$^-$, and C$_5$N$^-$\cite{Yen:JCPA:10}. The electronic                     
  absorption spectra  have been reported for  C$_7$N$^-$ and longer                
  cyanopolyyne anions in neon matrices\cite{Grutter:JCP:99}. The authors assigned the
  observed bands, progressively shifting to lower energies with the increasing carbon chain
  length,  to the $^1\Sigma^+ \leftarrow X^1\Sigma^+$ transition. This assignment was
  primarily  based on the analogy to the spectra of closed-shell neutral           
  polyacetylenes (HC$_{2n}$H).
  
  This paper reports  the calculations of the electronically  excited bound and    
  metastable states in                                                             
  cyanopolyyne anions  C$_{2n-1}$N$^-\; (n=1,\dots, 4)$.                           
  We employed the EOM-CCSD method\cite{Krylov:EOMRev:07,Christiansen:EOMRev:11,Bartlet:EOMREV:12} augmented by
  complex absorbing potential (CAP) to capture metastable states\cite{KrylovResReview}.
  EOM-CCSD  has shown robust performance in calculations of excited states of      
  different nature in closed-shell and open-shell species\cite{Krylov:EOMRev:07,Christiansen:EOMRev:11,Bartlet:EOMREV:12,RydbergReview09,Krylov:OSRev}.
  Recent implementation of the complex-valued CAP-EOM-CCSD                         
  variant\cite{Zuev:CAP:14,Jagau:CAP:13,Jagau:PES:14,Pal:CAPEOMCC:12,Sommerfeld:CAP-SACCI:12}
  extended this method to electronic resonances.                                   
  We find that C$_{2n-1}$N$^-$ possess  a distinct series of                      
  low-lying discrete excited states of singlet and triplet spin symmetry,          
  all classifiable  as valence states. These states appear already                 
  in CN$^-$ as resonances, i.e., above the lowest electron detachment threshold,   
  and then become  gradually  stabilized, such that they turn into                 
  the bound states as the length of the carbon chain increases.                    
  Energies of the transitions that                                                 
  correspond to $\pi \to \pi^\ast$ excitations can be  rationalized within the    
  H\"{u}ckel model, which predicts                                                 
  lowering of the excited states  in the molecules with the increasing             
  number of $\pi-$conjugated bonds.                                                
  Thus, the simple  H\"{u}ckel model retains its                                   
  semi-quantitative validity even for excited states of resonance character.       
  We analyze all identified excited states in terms of their lifetime (for the continuum
  part of the spectrum),                                                           
  orbital excitation character, Dyson orbitals, and transition dipole moments for  
  optically bright states.                                                         
  The large dipole moment of the lowest $^2\Sigma^+$ state in some of cyanopolyyne radicals
  suggest that they should also support dipole stabilized states. Indeed, we show  
  that C$_3$N$^-$ has dipole bound states while longer anions                      
  exhibit states which can be classified as dipole-stabilized resonances\cite{Jagau:AgCuF:15}.
  Finally, we discuss how the above features of electronic structure               
  can manifest themselves in experimental observables such as                      
  cross sections for photodetachment (from anions) or electron scattering (from neutrals)
  and comment on the implications of our results regarding                         
  the   mechanism  of  cyanopolyyne anions formation in the ISM.
  
  \section{Theoretical methods and computational details}                          
                                                                                   
\begin{figure}[h!]                                                               
  \includegraphics[width=8cm]{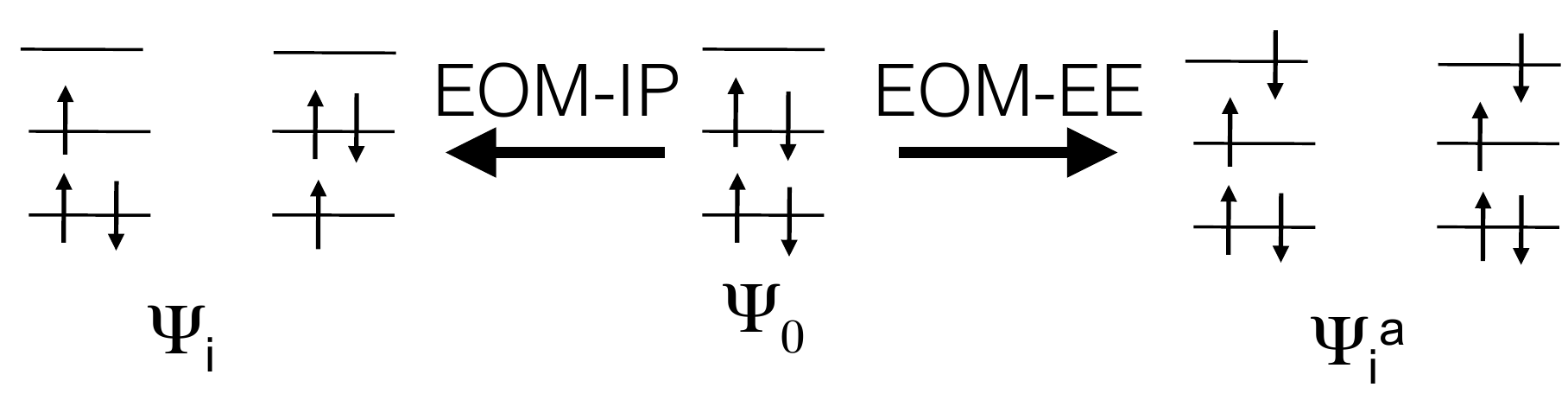}                                            
  \caption{                                                                        
    Manifolds of target states accessed by EOM-IP and EOM-EE. Here, $\Psi_0$ denotes
    closed-shell CCSD wave function of the ground-state anions. The two groups of target
    states are described by diagonalizing the same model Hamiltonian $\bar H$ obtained by the
    similarity transformation using the reference CCSD amplitudes.                 
  \label{fig:scheme}}                                                              
  \end{figure}

  We investigate the electronically excited states of the first four cyanopolyyne anions,
  C$_{2n-1}$N$^-$,                                                                 
  both below and above the electronic continuum onset, by means                    
  of the EOM-CCSD method\cite{Krylov:EOMRev:07,Bartlett:CC_review:07,Christiansen:EOMRev:11,Bartlet:EOMREV:12}.
  In all calculations, the reference wave function                                 
  corresponds to  the closed-shell ground-state CCSD solution for the  respective anion
  with $N$ electrons.                                                              
  We compute  excitation energies                                                  
  by the EOM-EE variant of the method in which $N$-electron target states are described by
  particle-conserving excitation operators. To describe $N$-electron metastable states
  embedded in the electron-detachment continuum, we augment                        
  the standard EOM Hamiltonian by complex absorbing potential                      
  (CAP-EOM-EE-CCSD)\cite{Zuev:CAP:14,Jagau:CAP:13,Jagau:PES:14}.                   
  We determine electron-detachment thresholds and properties of the parent neutral 
  radicals  by using  the EOM-IP                                                   
  variant of the method in which target $N-1$-electron states are described by EOM operators
  that remove one electron (i.e., 1 hole and 2-holes-one-particle).                
  Fig.~\ref{fig:scheme} illustrates                                                
  target-states manifolds accessed by  EOM-EE and EOM-IP.                          
                                                                                   
  Given our focus on metastable                                                    
  auto-ionizing states, we would like to highlight the advantages of the           
  EOM-CCSD approach for treating resonances\cite{KrylovResReview}. EOM-CC          
  is a  robust theoretical framework capable of treating diverse types of          
  electronic structure. The method yields size-intensive transition energies       
  (excitation, ionization, attachment) and their accuracy                          
  can be systematically improved towards the exact solution.                       
  In the context of metastable and near-threshold bound states, one of the         
  major advantages of the EOM-CC  approach  is that it                             
  provides a balanced description of different target states as a result of        
  employing the same model Hamiltonian in different EOM variants.                  
  Hence, the onsets of the                                                         
  continua in the EOM-EE calculations correspond exactly to the  electron detachment
  energies obtained in the EOM-IP calculations. Consequently, energy differences   
  are consistent with each other and one can unambiguously distinguish  between    
  the bound and continuum states.
  
  Use of CAP allows one to represent metastable states with $\mathcal{L}^2$-integrable    
  basis functions\cite{Meyer:CAP:93,Santra:CAPREV:02,Muga:CAP:04}.                 
  In the CAP calculations, regular Hamiltonian $\hat{H}_0$ is augmented by an imaginary
  potential $-i\eta \hat{W}$:                                                      
  \begin{equation}                                                                 
  \hat{H}(\eta)=\hat{H_0} -i \eta \hat{W}(r)                                       
  \end{equation}                                                                   
  where $\eta$ is the strength of the CAP. This purely imaginary potential         
  absorbs the diverging tail of the continuum states and transforms them into      
  $\mathcal{L}^2$-integrable wave functions with complex energies\cite{Sommerfeld:CSRev:02},
  which can be computed by standard {\it ab initio}                                
  approaches designed for bound states\cite{Cederbaum:csADC:02,Sommerfeld:CAPCI:01,Ernzerhof:CAPDFT:12,Pal:CAPEOMCC:12,Zuev:CAP:14,Jagau:CAP:13}.
  The CAP technique  is related to exterior complex scaling                        
  method\cite{Meyer:CAP:98}.                                                       
  In the present  study, we use the CAP-EOM-CCSD implementation                    
  reported in Refs. \citenum{Zuev:CAP:14,Jagau:CAP:13,Jagau:PES:14}. Our        
  calculations closely follow the protocol presented in those papers. The          
  CAP is introduced                                                                
  as a quadratic potential of a cuboid shape:                                        
  \begin{eqnarray}                                                                 
  \hat{W}(r)&=&\hat{W}_x(r_x)+\hat{W}_y(r_y)+\hat{W}_z(r_z), \nonumber\\
  \hat{W}_{\alpha}(r) &=& \begin{cases}
  0                           & \text{if} \; |r_{\alpha}| < r^0_{\alpha}\\       
  (|r_{\alpha}|-r^0_{\alpha})^2 & \text{if} \; |r_{\alpha}| > r^0_{\alpha},      
  \end{cases}                                                                      
  \end{eqnarray}                                                                   
  where the coordinates $(r_x^0, r_y^0, r_z^0)$ define the onset of CAP in each    
  direction. Following the same strategy as in previous calculations\cite{Zuev:CAP:14,Jagau:CAP:13},
  we fixed  the CAP onset                                                          
  at spatial extent of the wave function for the reference state,                  
  $r_i^0= \sqrt{\langle \Psi_{CCSD}|R_i^2|\Psi_{CCSD}\rangle}$,                    
  where $\Psi_{CCSD}$ is the CCSD solution (CAP-free) for the ground state of the anion\cite{CAPonset}.
  The CAP-augmented Hamiltonian yields complex eigenvalues $E(\eta) =E_R(\eta)-i\Gamma(\eta)/2$
  where $E_R$  is the position of the resonance and $\Gamma(\eta)$ is its          
  width.                                                                           
  The optimal value of the CAP strength  parameter $\eta$ is determined            
  for each metastable state by calculating $\eta$-trajectories and searching for   
  the minimum  of the function\cite{Meyer:CAP:93}:                                 
  \begin{equation}                                                                 
  |\eta \cdot\frac{d E(\eta) }{d \eta}|= \text{min},                                      
  \label{eq:vel}                                                                   
  \end{equation}                                                                   
  which minimizes the error introduced by the incompleteness of the                
  one-electron basis set and finite strength of CAP.                               
  In some cases, more robust and stable resonance parameters can be obtained if,   
  instead of considering the raw trajectories $E(\eta)$,                           
  one analyzes the $\eta$-trajectories of deperturbed energies ($U(\eta)$) from which
  the explicit dependence on CAP is removed in the first                           
  order\cite{Jagau:CAP:13,Jagau:PES:14}.                                           
  The deperturbed complex energies $U(\eta)$  are calculated by                    
  subtracting from raw energies $E(\eta)$  the correction $i\eta                   
  \;\textrm{Tr}[\gamma \;W]$ where $\gamma$ is the one-particle density matrix     
  of the resonance state.
  
  Characters of the excited states                                                 
  can be illuminated by  Dyson orbitals and transition dipole                      
  moments\cite{Jagau:Dyson:16}. We also used EOM-CC  Dyson orbitals to compute     
  the photoelectron dipole matrix element needed for calculating                   
  of photodetachment cross section from anionic species.                           
  In these                                                                         
  calculations the outgoing electron was represented by plane waves, similarly     
  to our previous studies\cite{Gozem:Dyson:15}.                                    
                                                                                   
  Unless otherwise specified, in  excited-state calculations                       
  we employ the aug-cc-pVTZ  basis set\cite{Dunning:92:augccpvxz} augmented        
  with an extra set of diffuse function (mostly $3s3p$) centered on each atom.     
  The exponents of these additional functions were obtained                        
  in the same way as in our previous studies\cite{Zuev:CAP:14,Jagau:CAP:13,Jagau:PES:14}.
  The calculations were performed at the fixed                                     
  equilibrium geometries of the respective anions.                                 
  The optimal geometries of the anions were taken from the                         
  literature. They correspond either to the experimental values                    
  (CN$^-$)\cite{Bradforth:JCP:98}                                                  
  or highly accurate theoretical optimization at the CCSD(T)/aug-cc-pVQZ level     
  [(C$_3$N$^-$)\cite{Stein:MP:15},                                                 
  (C$_5$N$^-$)\cite{Botschwina:JCP:08}, (C$_7$N$^-$)\cite{Botschwina:JCP:08}].     
  The Cartesian coordinates  are given  in the Supplementary Information.          
  In calculations of the adiabatic detachment energies, we optimized               
  the structures of the neutral precursors by  EOM-IP-CCSD/cc-pVTZ.                
  The electronic structure  calculations reported here were performed  using the   
  Q-Chem  package\cite{qchem_2014s,qchem_feature}. 
  
  \section{Results}

\begin{figure}[h]
  \centering
    \includegraphics[width=5.5cm]{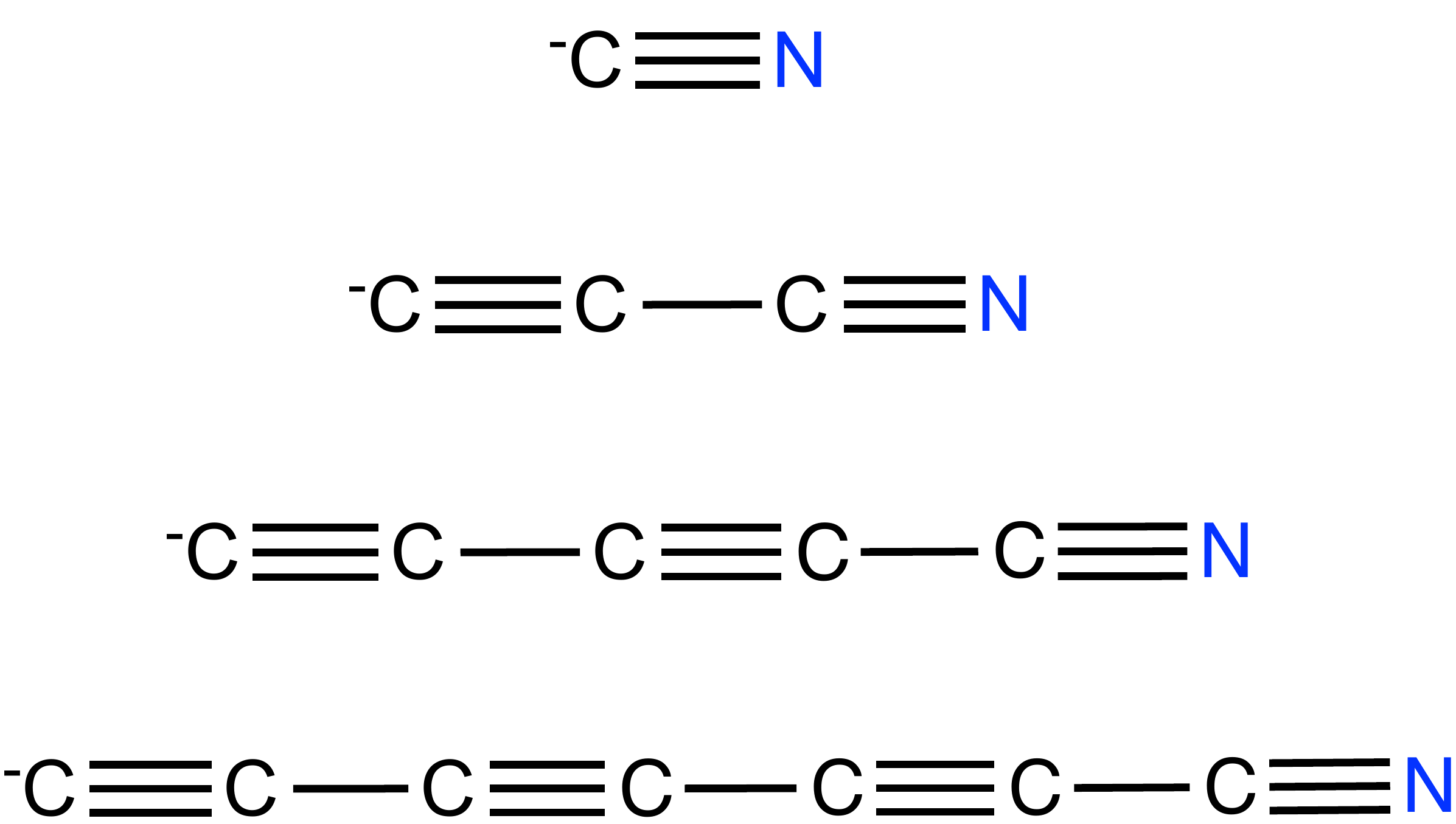}                                    
    \caption{                                                                      
     Four cyanopolyyne anions studied in this work.                                
      \label{fig:mol}}                                                             
  \end{figure}

\begin{figure}[!b]
  \centering
    \includegraphics[width=8cm]{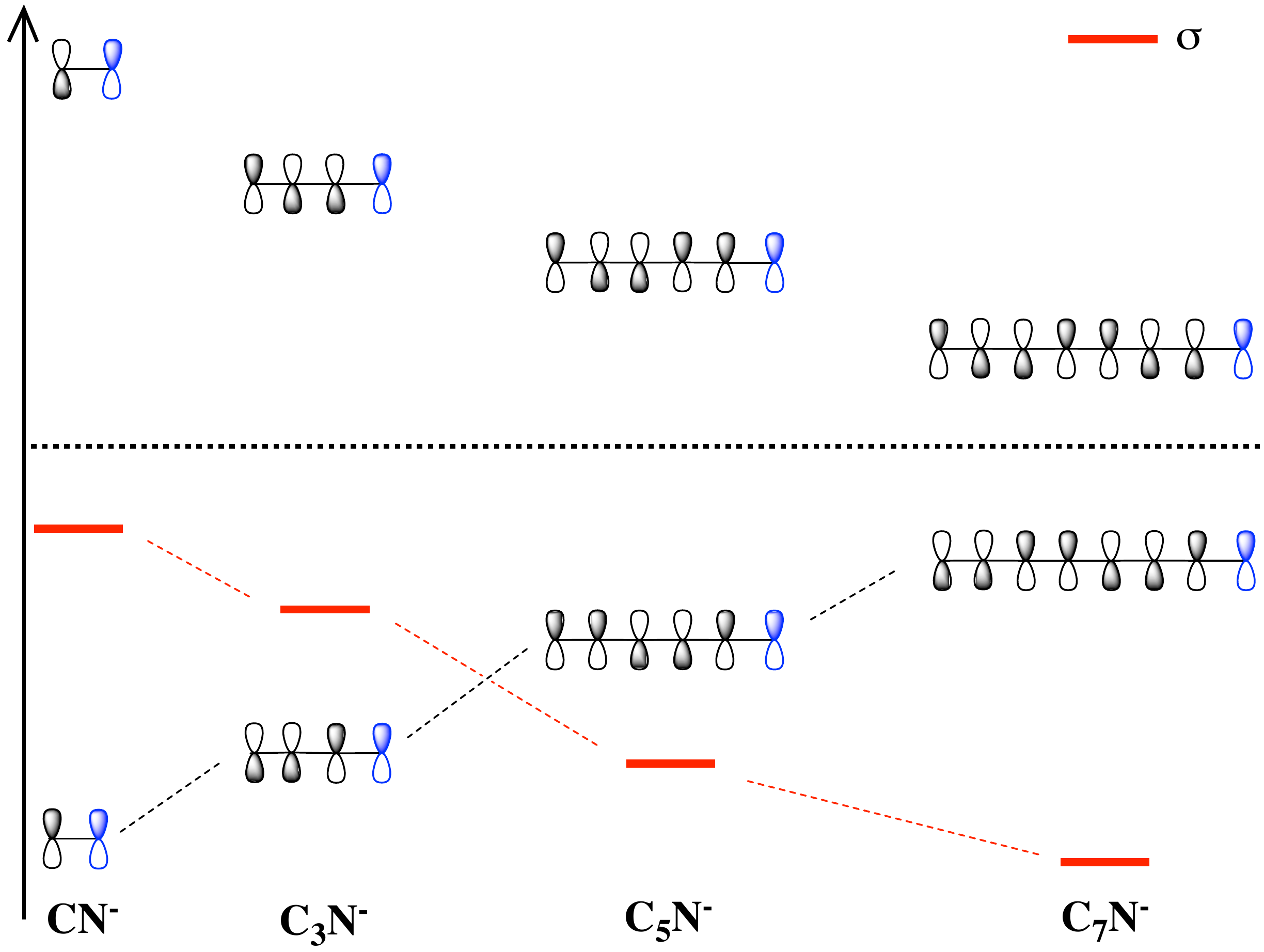}                                 
    \caption{                                                                      
     Frontier molecular orbitals in cyanopolyyne anions.                           
      \label{fig:piorb}}                                                           
  \end{figure}

  All cyanopolyyne anions                                                          
  considered in this work are                                                      
  linear species with alternating single and triple bonds (Fig.~\ref{fig:mol}).    
  In the ground state, they                                                        
  have a closed-shell                                                              
  electronic configuration of $^1\Sigma^+$ symmetry.                               
  The molecular orbital analysis of the detached and excited states                
  studied here                                                                     
  reveals that two classes of orbitals are involved in                             
  the relevant electronic transitions. 
  The first class comprises                   
  the two sets of orthogonal and degenerate $\pi$ orbitals. The second             
  class comprises lone pair located on the terminal carbon  --- this orbital is    
  of a $\sigma$ type.                                                              
  The $\pi$ system in the C$_{2n-1}$N$^-$ anions is similar to the $\pi$ system in conjugated
  polyynes. Hence, one can employ the                                              
  H\"{u}ckel model to interpret the trend in energies                            
  of the $\pi$ orbitals with respect to the carbon chain length.                   
  In accordance with the H\"{u}ckel model predictions, the energy of the           
  frontier $\pi$ orbitals increases (as does  the respective detachment energy) and
  the energy of $\pi^*$ orbitals decreases, as the length of the carbon chain increases. This
  trend is illustrated in Fig.~\ref{fig:piorb}.                                    
  In contrast, the $\sigma$ orbital  becomes more bound in longer carbon chains,   
  which can be rationalized in terms of electrostatic interaction between the lone 
  electron pair and the increasing dipole moment of the  neutral core.

\begin{figure}[h!]
    \centering
    \includegraphics[width=7cm]{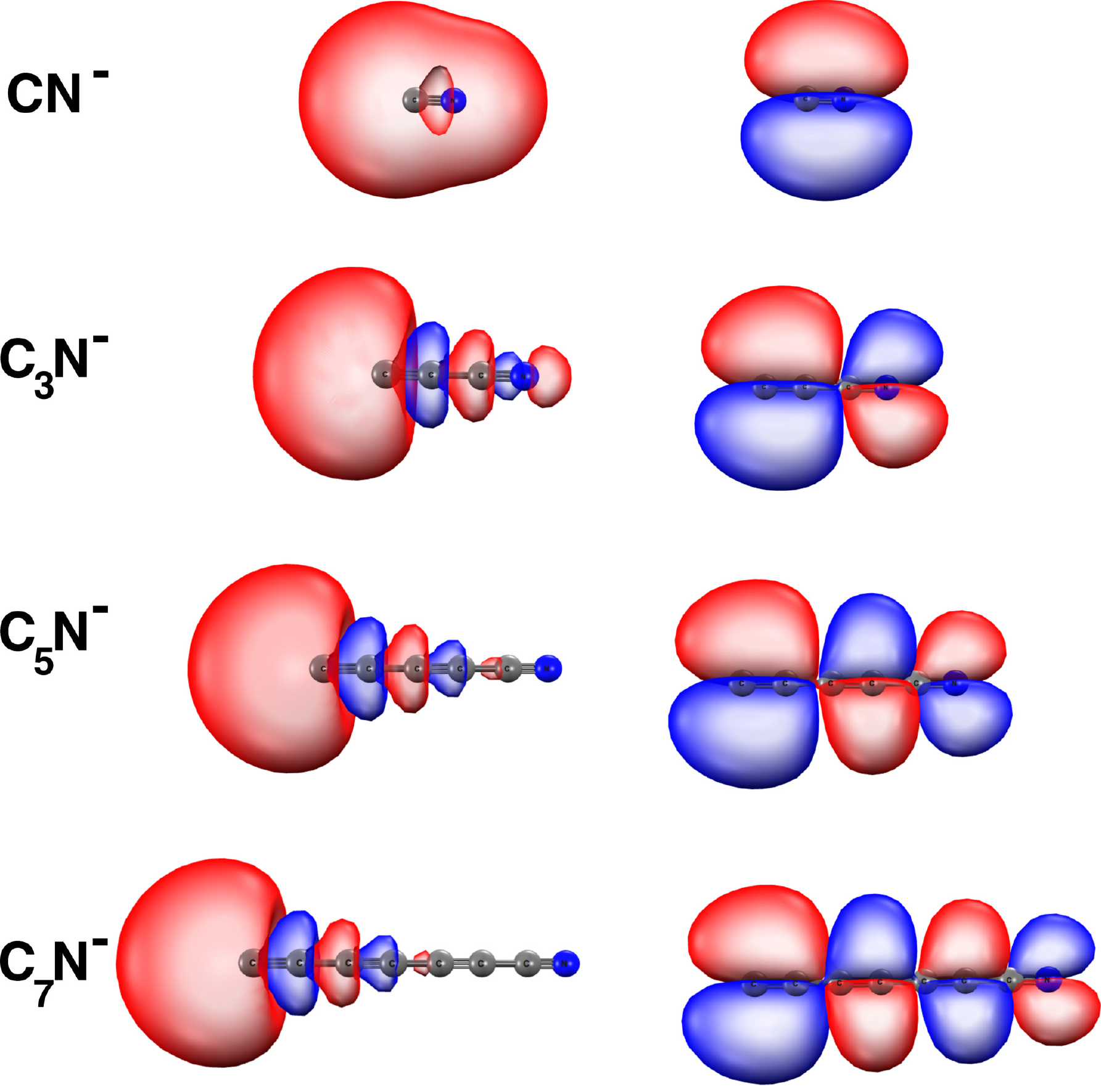}                                         
    \caption{                                                                      
     Dyson orbitals of  cyanopolyyne anions corresponding to formation             
     of the $^2\Sigma^+$ (left) or $^2\Pi$ (right) neutral radicals.                 
    \label{fig:dyson}}                                                             
  \end{figure}

The  opposite trends in energetics  of the frontier  occupied orbitals ($\pi$ and $\sigma$)
  clearly manifest themselves in the computed electron detachment                  
  energies. Table~\ref{tab:vde} presents  calculated  vertical                     
  detachment energies for the  two lowest detached states  of $^2\Sigma^+$  and    
  $^2\Pi$ symmetry.  A $^2\Sigma^+$ radical is obtained by removing an electron    
  from the $\sigma$ orbital of the anion, whereas electron detachment from $\pi$   
  orbital  leads to a $^2\Pi$ radical, as  illustrated by the                      
  shapes of the corresponding Dyson orbitals in Fig.~\ref{fig:dyson}.

  \begin{table}                                                                
  \renewcommand{\arraystretch}{1.2}                                                
  \setlength{\tabcolsep}{9pt}                                                      
    \caption{Vertical detachment energies (VDE) of the anions (in eV) and dipole   
  moments $\mu$ (in debye) of the neutral                                          
  radicals.                                                                        
  \label{tab:vde}}                                                                 
  \begin{tabular}{ccccc}                                                           
  \hline                                                                           
  Molecule  &  VDE ($^2\Sigma^+$)  &  $\mu(^2\Sigma^+)$  & VDE ($^2\Pi$) &  $\mu (^2\Pi)$    \\
  \hline                                                                           
  CN$^-$                &   3.99   &   1.35   &   5.28   &   0.18   \\             
  C$_3$N$^-$            &   4.67   &   3.87   &   4.79   &   0.14   \\             
  C$_5$N$^-$            &   4.98   &   5.86   &   4.70   &   0.11   \\             
  C$_7$N$^-$            &   5.22   &   7.97   &   4.71   &   0.03   \\             
  \hline                                                                           
  \end{tabular}                                                                    
  \end{table}

  \begin{figure}[h!]
  \centering
  \includegraphics[width=6cm]{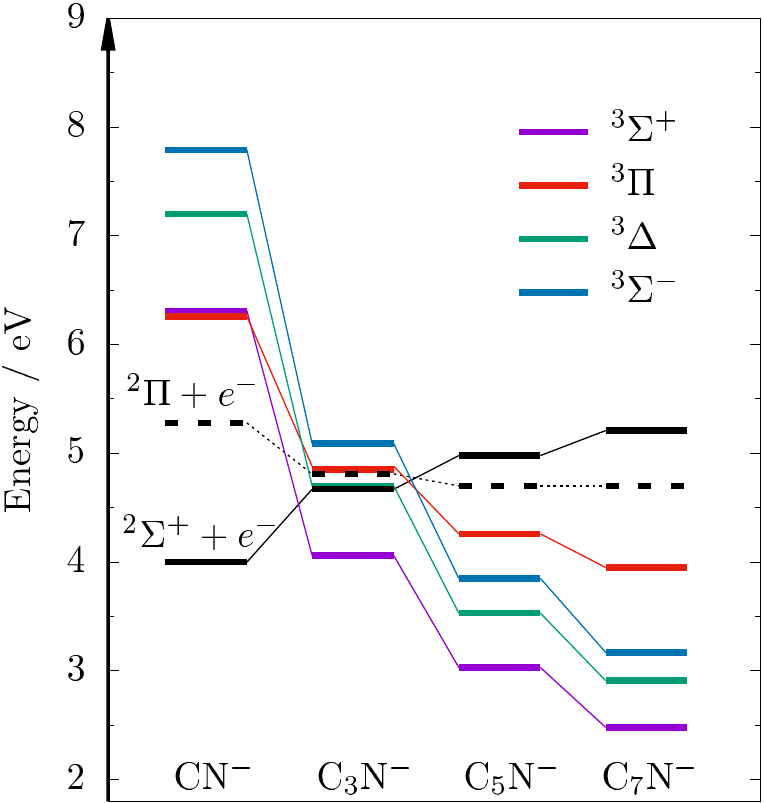}\\ 
  \vspace{0.5cm}
  \includegraphics[width=6cm]{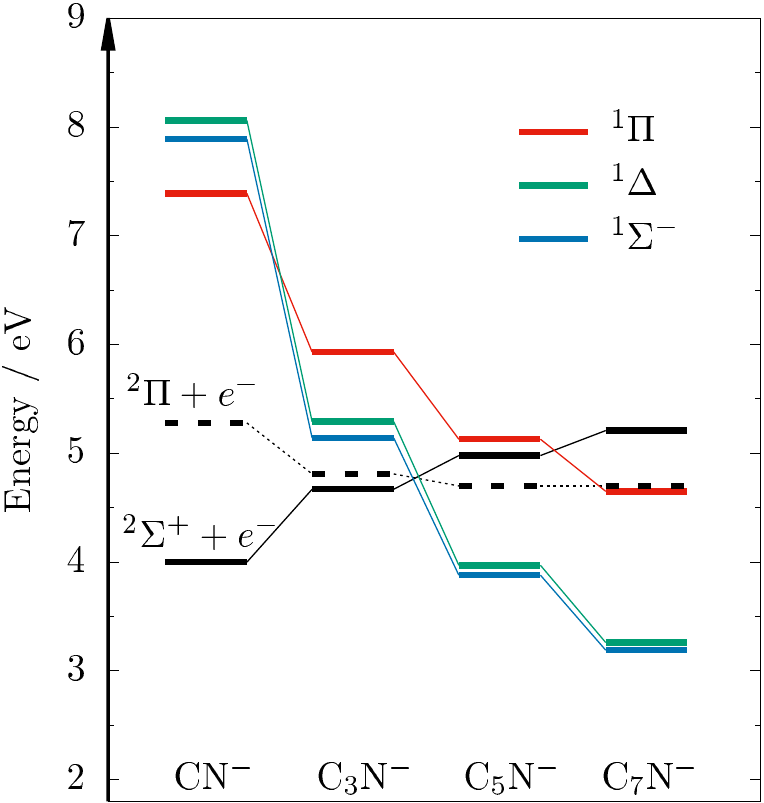}                                        
  \caption{                                                                      
    Valence excited states in cyanopolyyne anions. Top                            
    panel: triplet states, bottom panel: singlet states. Black                      
    lines mark the two lowest electron detachment thresholds corresponding to      
    formation of $^2\Sigma^+$ (solid) or $^2\Pi$ (dashed) neutral radicals.        
    \label{fig:endiag}                                                             
    }                                                                              
  \end{figure}

  As expected, the Dyson                                                           
  orbitals for the $^2\Sigma^+$ states have axial symmetry with highest            
  electron density at the carbon end of the chain.                                 
  The  Dyson orbitals for the $^2\Pi$ states have a typical nodal                  
  structure of $\pi$ orbitals                                                      
  and are delocalized over the entire chain, following the H\"{u}ckel model        
  prediction (Fig. \ref{fig:piorb}).
Table~\ref{tab:vde} shows that all anions have                                   
  large electron detachment energy:                                                
  $\sim4\;$eV or more. Forming $^2\Sigma^+$ radicals is more favorable in          
  shorter carbon chains, whereas                                                   
  $^2\Pi$ radicals become preferable in longer species, which leads to             
  the change of the                                                                
  ground state from $^2\Sigma^+$ to $^2\Pi$ as the carbon chain increases.   
  Directly comparable with experimental measurements  are adiabatic detachment energies
  (ADE). Table~\ref{tab:ade} summarizes our calculated ADEs together with the available  experimental values\cite{Bradforth:JCP:98,Yen:JCPA:10,Graupner:NJP:06}.
 The discrepancies between the theory and experiment                              
  are within  0.2--0.3 eV, as expected for EOM-CCSD.                               
  According to our calculations,  the cross-over between the                       
  $^2\Sigma^+$ and $^2\Pi$  states                                                 
  occurs                                                                           
  in C$_5$N, both adiabatically and vertically.                                    
  The most recent photoelectron spectroscopic study\cite{Yen:JCPA:10}              
  found that the  ground state of C$_5$N is still $^2\Sigma^+$,                    
  whereas  the $^2\Pi$ state is located $0.069\pm 0.015\;$eV above the $^2\Sigma^+$
  threshold. Thus,                                                                 
  a higher level of electron correlation treatment beyond EOM-CCSD is necessary to 
  exactly reproduce                                                                
  the cross-over between these two nearly degenerate states of C$_5$N.
    \begin{table}[b!]                                                                
  \renewcommand{\arraystretch}{1.2}                                                
  \setlength{\tabcolsep}{9pt}                                                      
    \caption{Adiabatic detachment energies (in eV). Theoretical values neglect     
    zero point energy contributions.                                         
    \label{tab:ade}}                                                               
  \begin{tabular*}{0.48\textwidth}{@{\extracolsep{\fill}}lll}
  \hline                                                                           
   Molecule  &  $^2\Sigma^+$ &   $^2\Pi$   \\                                      
  \hline                                                                           
  CN$^-$ (this work)                &   3.99                  &   5.16   \\        
  Exp. (Ref. \citenum{Bradforth:JCP:98}) &  $3.862 \pm 0.004$      &          \\
  \hline                                                                           
  C$_3$N$^-$ (this work)            &   4.54                  &   4.75    \\       
  Exp. (Ref. \citenum{Yen:JCPA:10})      &  $4.305 \pm  0.001$     &          \\
  Exp. (Ref. \citenum{Graupner:NJP:06})  &  $4.59  \pm  0.25 $     &          \\
  \hline                                                                           
  C$_5$N$^-$ (this work)            &   4.79                  &   4.69   \\        
  Exp. (Ref. \citenum{Yen:JCPA:10})      &  $4.45  \pm  0.03$      &                  \\
  \hline                                                                           
  \end{tabular*}                                                                    
  \end{table}                                                                       
                                                                                    
  An important property  of  the C$_{2n-1}$N radicals is  their dipole moment,     
  which is  given in Table~\ref{tab:vde} for the two lowest electronic states.     
  The states of  $^2\Sigma^+$ symmetry  have large dipole                          
  moment,                                                                          
  which increases with the length of the carbon chain.                             
  The $^2\Pi$  states behave differently -- their dipole moment                    
  remains small, independently  of the molecular size.                             
  These trends in the dipole moments have two                                      
  consequences. First, it explains the stabilization of the lone pair              
  orbital, as seen in the energies of the $^2\Sigma^+$ detached states.            
  Second, it means that the neutral C$_{2n-1}$N  radicals should be capable of forming
  dipole-bound states\cite{Simons:08:MolAnions,Jordan:DBRev:03} (or dipole-stabilized
  resonances\cite{Jagau:AgCuF:15})  only if they are in the                        
  $^2\Sigma^+$ state, but not in  the $^2\Pi$ state (in other words,               
  dipole-stabilized states should correspond to the excitation of an electron from the
  $\sigma$ orbital to a diffuse target orbital near the carbon end).

  The  electron detachment energies determine the thresholds that separate         
  bound and unbound (metastable) excited states. The valence excited               
  states of cyanopolyyne anions studied in the present work are predominantly the  
  excitations from the frontier $\sigma$ or $\pi$  orbital into the unoccupied $\pi^\ast$
  orbital. The $\sigma \to \pi^\ast$  transition may result in singlet or triplet $\Pi$
  states, whereas the $\pi \to \pi^\ast$  transition can give rise to singlet or
  triplet $\Sigma^+$, $\Sigma^-$,  or  $\Delta$
  states. Several possible excited states symmetries                       
  result from double degeneracy of  $\pi-$system.                                  
  Fig.~\ref{fig:endiag} shows energies of all low-lying valence excited states,  
  either metastable or bound.                                                      
  These energy diagrams illustrate that all                                        
  excited states of $\sigma \to \pi^\ast$ and  $\pi \to \pi^\ast$                
  character                                                                        
  become stabilized with the increasing length of the carbon chain.                
  The rate of stabilization is  different in the excited stated derived from       
  the $\pi \to \pi^\ast$ transitions and in those derived by the  $\sigma \to \pi^\ast$
  excitations.                                                                     
  The trend in the energies of all                                                 
   $\pi \to \pi^\ast$  transitions can be rationalized in terms of the H\"{u}ckel model.
  The  $\sigma \to \pi^\ast$ transitions                                          
  follow a different pattern,  determined by the opposing trends in the $\pi^\ast$
  and $\sigma$ orbitals.                                                           
  More thorough  analysis of these trends in electronic spectra                    
  is given in Section \ref{sec:GeneralTrends},                                
  after detailed                                                        
  results for each species, starting from   CN$^-$, the                            
  prototype of  all cyanopolyyne anions.
 
    \begin{table*}                                                               
  \renewcommand{\arraystretch}{1.2}                                                
  \setlength{\tabcolsep}{9pt}                                                      
  \caption{Positions $E_R$, widths $\Gamma$ (eV), optimal values                   
  of $\eta$ parameter, and corresponding trajectory velocities (in atomic          
  units) for low-lying resonances in CN$^-$.                                       
  Basis set is aug-cc-pVTZ+3s3p.                                                   
  Numbers in parentheses denote powers of 10.                                      
  \label{tab:cn1}}                                                                 
  \begin{tabular*}{\textwidth}{@{\extracolsep{\fill}}lllllllll}                                                       
  \hline                                                                           
  State& $E_R^{(0)}$  & $\Gamma^{(0)}$  &                                          
  $\eta_{\textrm{opt}}^{(0)}$  & $v^{(0)}$  &                                      
  $E_R^{(1)}$  & $\Gamma^{(1)}$                                                    
  & $\eta_{\textrm{opt}}^{(1)} $  & $v^{(1)}$                                      
  \\                                                                               
  \hline                                                                           
  $^3\Sigma^+  (\pi \to \pi^\ast)$    &  6.34     &   0.20    &  0.0034  &  1.2(-3)   &    6.31    &    0.16     &  0.007 &  5.9(-4)   \\
  $^3\Pi       (\sigma \to \pi^\ast)$ &  6.62     &   0.61    &  0.055  &   4.5(-3)   &    6.56    &    0.45     &  0.12  &  3.0(-3)   \\
  $^3\Delta    (\pi \to \pi^\ast)$    &  7.51     &   0.58    &  0.055  &   4.7(-3)   &    7.45    &    0.42     &  0.10  &  2.1(-3)   \\
  $^3\Sigma^-  (\pi \to \pi^\ast)$    &  8.18     &   0.64    &  0.055  &   4.4(-3)   &    8.19    &    0.48     &  0.12  &  2.1(-3)   \\
  $^1\Pi       (\sigma \to \pi^\ast)$ &  7.90     &   0.79    &  0.064  &   3.7(-3)   &    7.88    &    0.56     &  0.18  &  2.0(-3)   \\
  $^1\Delta    (\pi \to \pi^\ast)$    &  8.54     &   0.70    &  0.056  &   4.5(-3)   &    8.49    &    0.52     &  0.14  &  1.5(-3)   \\
  $^1\Sigma^-  (\pi \to \pi^\ast)$    &  8.29     &   0.65    &  0.052  &   4.4(-3)   &    8.23    &    0.48     &  0.12  &  2.0(-3)   \\                                                                          
  \hline                                                                           
  \end{tabular*}                                                                    
  \end{table*}
  \subsection{CN$^-$}                                                              
  CN$^-$ is the precursor of the cyanopolyynes series. Owing to its small size,    
  it often serves as a benchmark                                                   
  system amenable to high-level quantum-chemical calculations.                     
  CN$^-$  and its parent radical CN  play important roles  not only in the ISM     
  and astrochemistry, but also in combustion chemistry, carbon-rich plasmas, rotational
  spectroscopy, and in precision measurements                                      
  of cosmological properties\cite{Gottlieb:JCP:07,Leach:MNRAS:12,Casavecchia:FD:02}.
  Our CAP-EOM-EE-CCSD calculations                                                 
  identified  seven valence low-lying resonances of different                      
  symmetries. Their positions  and linewidths  are reported in Table~\ref{tab:cn1} and
  the $\eta$ trajectories used to extract the resonance                            
  parameters are given in the Supplementary Information.                           
  The energies are given with respect to the ground state of CN$^-$. The resonances
  are located                                                                      
  in the range of 6.3 to 8.5 eV above the ground state, that is,                   
  more than 2 eV above the lowest CN$^-$ detachment threshold.                     
  They are all predominantly $\pi \to \pi^\ast$ excitations, except for           
  the $^1\Pi$ and $^3\Pi$                                                          
  states  that correspond to $\sigma \to \pi^\ast$ transition.
   \begin{table}[!b]                                                                
  \renewcommand{\arraystretch}{1.2}                                                
  \caption{Positions $E_R$ (and widths $\Gamma$), in eV, of the  resonances in CN$^-$
    calculated with various basis sets.
  \label{tab:cn2}}                                                                 
  \begin{tabular*}{0.48\textwidth}{@{\extracolsep{\fill}}lllc}                                                                                                                \hline                                                                           
                &       aug-cc-pVTZ   &      aug-cc-pVQZ  &    aug-cc-pVTZ  \\
                &       +3s3p         &      +3s3p         &   +4s4p3d   \\
    \hline                                                                           
  $^3\Sigma^+$  &     6.34 (0.20)     &    6.35 (0.19)     &    6.31 (0.19)       \\
  $^3\Pi$       &     6.62 (0.61)     &    6.49 (0.49)     &    6.26 (0.51)       \\
  $^3\Delta$    &     7.51 (0.58)     &    7.40 (0.45)     &    7.20 (0.38)       \\
  $^3\Sigma^-$  &     8.18 (0.64)     &    8.03 (0.52)     &    7.79 (0.55)       \\
  $^1\Pi$       &     7.90 (0.79)     &    7.69 (0.74)     &    7.39 (1.05)       \\
  $^1\Delta$    &     8.54 (0.70)     &    8.37 (0.57)     &    8.06 (0.70)       \\
  $^1\Sigma^-$  &     8.29 (0.65)     &    8.15 (0.52)     &    7.89 (0.58)       \\
  \hline                                                                           
  \end{tabular*}                                                                    
  \end{table} 
  
  All states can decay to a parent radical ($^2\Sigma^+$ or $^2\Pi$)               
  in one-electron process, thus, they can be classified as shape resonances.       
  Inspection of Table~\ref{tab:cn1} shows                                          
  that all singlet resonances are slightly broader than the triplets of the same   
  symmetry and that they are higher in energy. The results obtained from           
  the uncorrected trajectories ($E_R^{(0)}$, $\Gamma^{(0)}$) and the first-order corrected
  trajectories  ($E_R^{(1)}$, $\Gamma^{(1)}$) differ by no more than 0.1 eV        
  for the position of the resonances and by no more than 0.25 eV for the           
  widths. The optimal velocities, Eq. (\ref{eq:vel}),                              
  obtained from the first-order corrected                                          
  trajectories $v^{(1)}$ are systematically lower than those from the  uncorrected 
  trajectories $v^{(0)}$,                                                          
  confirming  that the correction\cite{Jagau:CAP:13}                               
  indeed improves                                                                  
  the accuracy of the resonance determination.                                     
  The small size of cyanide allows us to                                       
  test the convergence of the results with respect to the size of the one-electron 
  basis set. This is summarized  in Table~\ref{tab:cn2}, which presents            
  the positions and widths                                                         
  of the CN$^-$ resonances obtained with three different basis sets.               
  Table~\ref{tab:cn2} shows that the smallest tested basis set    
  (aug-cc-pVTZ+3s3p) gives already satisfactory results for the  resonance         
  parameters, in particular, for the narrowest resonances. In most                 
  cases, larger valence  or more diffuse basis sets lead to slightly lower         
  resonance positions and smaller widths. Importantly, in all cases, the resonance 
  parameters obtained with our basic basis set (aug-cc-pVTZ+3s3p) are reasonably   
  close to the position and widths  obtained  with larger basis sets. On           
  the basis of these benchmarks, for larger cyanopolyyne anions we employ only     
  the aug-cc-pVTZ+3s3p basis set.
  
  To further validate our results from CAP calculations,                           
  we  computed the resonance parameters for                                        
  the CN$^-$ triplet states with the complex basis function (CBF) method\cite{Head-Gordon:CS:15,White:17b} (see Supplementary
  Information for the details). The results from the CAP                           
  (Table~\ref{tab:cn2})  and from the CBF (Table~SII) calculations                 
  compare favorably,                                                               
  reaffirming  our predictions of resonance states in  CN$^-$.                     
  We can also directly compare our results for CN$^-$ with the $R$-matrix calculations
  of Harrison and Tennyson\cite{Harrison:JPB:12}.                                  
  They reported  three                                                             
  triplet resonances ($^3\Sigma^+$, $^3\Pi$ and $^3\Sigma^-$), located 3.2 to 4.9  
  eV above the lowest electron detachment threshold with widths around $\sim 1$ eV 
  (see Table 3 in Ref.~\citenum{Harrison:JPB:12} for the                        
  exact numbers obtained with different                                            
  electronic structure models employed within the $R-$matrix approach).            
  Due to strong dependence of the R-matrix results on the electronic structure
  model, the
  authors estimated  the uncertainty in their reported widths up to 50\%. 
  When compared with our results, it appears that the method applied in Ref. \citenum{Harrison:JPB:12}          
  yields  larger resonance widths, which 
  led to the conclusion that singlet resonances are not supported by CN$^-$. 
  Several other studies computed excited states in  CN$^-$                         
  using methods for bound electronic states\cite{Ha:MP:80,Polak:JMST:02,Musial:MP:05}.
  Interestingly, potential energy curves for CN$^-$ from  Ref. \citenum{Polak:JMST:02} show that
  $^3\Sigma^+$ state becomes stable with respect to electron detachment at the bond length
  beyond $\sim$2.9 bohr. This prediction is in line with our CAP results showing   
  that the $^3\Sigma^+$ resonance is the most narrow one at the equilibrium structure.
  Another reason why the $^3\Sigma^+$ state is  potentially  the most              
  easily stabilized resonance is  that it is the only  state  that                 
  asymptotically correlates with the lowest atomic threshold:                      
  C$^-$($^4\textrm{S}_u$) + N($^4\textrm{S}_u$). This threshold is located about 1.26 eV  below
  the lowest C($^3\textrm{P}_g$) + N($^4\textrm{S}_u$) asymptote, meaning that sufficiently
  stretched CN$^-$  must  have stable (with respect to electron detachment)
  excited states of $^3\Sigma^+$, $^5\Sigma^+$, and $^7\Sigma^+$ symmetry.         
                                                                                   
  We can also relate our results for  metastable states of CN$^-$ to               
  the electronic structure of isoelectronic diatomic molecules, NO$^+$             
  (Ref. \citenum{Partridge:JCP:90})                                             
  and N$_2$ (Ref. \citenum{Ermler:JPC:82}). All predicted  CN$^-$               
  resonances follow closely the pattern of the low-lying excited states in NO$^+$ and N$_2$.
  For example, NO$^+$  possesses  a manifold of valence excited states             
  in the energy range of 6.3 -- 9.0 eV above the ground                            
  state, which have exactly the same symmetries and orbital characters             
  as the computed CN$^-$ resonances. Similarly to CN$^-$,  neither NO$^+$ nor N$_2$
  have low-lying excited states of the  $^1\Sigma^+$ symmetry that                 
  can be derived from the $\pi \to \pi^\ast$ transition. 
  
  \subsection{C$_3$N$^-$}
   \begin{table*}[h!]                                                                
  \renewcommand{\arraystretch}{1.2}                                                
  \caption{Positions $E_R$, widths $\Gamma$ (in eV), optimal values                
  of $\eta$ parameter, and corresponding trajectory velocities (in atomic          
  units) for low-lying resonances in C$_3$N$^-$. $^3\Sigma^+$ is                   
  a bound state. Basis set is aug-cc-pVTZ+3s3p.                                    
  Numbers in parentheses denote powers of 10.                                      
  \label{tab:c3n}}                                                                 
  \begin{tabular*}{\textwidth}{@{\extracolsep{\fill}}lllllllll}
  \hline                                                                           
  State & $E_R^{(0)}$  & $\Gamma^{(0)}$  &                                         
  $\eta_{\textrm{opt}}^{(0)}$  & $v^{(0)}$  &                                      
  $E_R^{(1)}$  & $\Gamma^{(1)}$                                                    
  & $\eta_{\textrm{opt}}^{(1)}$  & $v^{(1)}$                                       
  \\                                                                               
  \hline                                                                           
  $^3\Sigma^+  (\pi \to \pi^\ast)$    &  4.06     &   ---     &                     &                       &            &             &                     &                        \\
  $^3\Pi       (\sigma \to \pi^\ast)$ &  4.85     &   0.015   &  0.00065 &   1.2(-4)   &    4.85    &    0.011    &  0.0016  &  5.0(-5)     \\
  $^3\Delta    (\pi \to \pi^\ast)$    &  4.68     & $<0.001$  &          &         &    4.68    &             &       &          \\
  $^3\Sigma^-  (\pi \to \pi^\ast)$    &  5.10     &   0.017   &  0.0010  &   1.4(-4)   &    5.10    &    0.012    &  0.0018  &  8.7(-5)    \\
  $^1\Pi       (\sigma \to \pi^\ast)$ &  5.94     &   0.152   &  0.0085  &   6.6(-4)   &    5.92    &    0.123    &  0.0150   &  1.5(-4)     \\
  $^1\Delta    (\pi \to \pi^\ast)$    &  5.30     &   0.029   &  0.0010  &   1.7(-4)   &    5.29    &    0.025    &  0.0020  &  1.8(-4)     \\
  $^1\Sigma^-  (\pi \to \pi^\ast)$    &  5.16     &   0.022   &  0.0015  &   1.7(-4)   &    5.16    &    0.017    &  0.0030  &  1.5(-4)     \\
  \hline                                                                           
  \end{tabular*}                                                                    
  \end{table*} 
  Results of the CAP-EOM-CCSD calculations for C$_3$N$^-$,                         
  the second cyanopolyyne anion,  are presented in Table~\ref{tab:c3n}.            
  The low-lying resonances of  C$_3$N$^-$ appear                                   
  at lower energies relative to CN$^-$. The $^3\Sigma^+$                           
  state is below the lowest detachment threshold and, therefore, is bound          
  with respect to electron detachment.                                             
  While being red-shifted, all C$_3$N$^-$ resonances                               
  also become more narrow by roughly one order of magnitude relative to CN$^-$.    
  Table~\ref{tab:c3n} shows that the observed stabilization effect is accompanied by
  much lower optimal values of the CAP strength parameter $\eta$ and much lower    
  corresponding trajectory velocities. 
  This means that determination of resonances parameters for C$_3$N$^-$            
  should be more robust and less basis set dependent than for CN$^-$.              
  While the $^3\Sigma^+$ state became a valence bound state, the                   
  $^3\Delta$  state  changed its character from shape to Feshbach resonance, i.e., 
  the $^3\Delta$ state                                                             
  (of $\sigma^2 \pi^2 \pi \pi^{\star}$ electronic configuration)                   
  is located below its parent neutral state of the C$_3$N radical                  
  ($^2\Pi$, electron configuration $\sigma^2 \pi^2  \pi$) but can still            
  autoionize producing the $^2\Sigma$ state of C$_3$N (of $\sigma \pi^2 \pi^2$ electronic
  configuration) via two-electron transition. This change in the resonance character
  leads to longer lifetime and ultra-narrow resonance width,                       
  lower than the accuracy of our method. This is why for the  $^3\Delta$ resonance we
  were only able to obtain an upper bound of the width.                            
                                                                                   
  Similarly to CN$^-$, the metastable electronic states of C$_3$N$^-$ were         
  investigated by the $R-$matrix method\cite{Harrison:JPB:11}.                
  Comparison of our results with those                                             
  given in Table 7 of Ref. \citenum{Harrison:JPB:11} shows  qualitative similarity
  between the CAP-EOM-EE-CCSD and R-matrix resonance parameters. In both calculations,
  the order of all resonance states  is  the same and                              
  the C$_3$N$^-$ resonances  are                                                   
  significantly more narrow than the corresponding states in CN$^-$.               
  The most notable                                                                 
  difference is in the widths, which are a few times larger in                     
  the R-matrix calculations. Moreover,                                             
  the resonances in the R-matrix calculations appear                               
  about  1.5 eV higher than in the CAP-EOM-CCSD calculations.                      
  Slightly  different geometries employed  in the two studies                      
  would not account for the   discrepancies in resonance                           
  predictions; rather, those  discrepancies arise due to                           
  the differences in the description of the continuum  
  and its coupling to the bound  part of the
  spectrum as well as approximations in the many-body treatment.
  In principle, the hermitian R-matrix scattering
  approach is well suited for treating resonances.
  However, the many-body nature of the molecular electronic problem implies 
  that other factors, i.e., electron correlation treatment and one-electron
  basis sets, contribute to the observed discrepancies.                                        
  This is confirmed by significant discrepancies across different electronic structure 
  models employed in the R-matrix study, as can be seen  in Table 7 of Ref.
  \citenum{Harrison:JPB:11}.
  A systematic comparison of the  two approaches (CAP-EOM-CCSD and R-matrix) would be very
  meaningful in this context.
  
  \subsection{C$_5$N$^-$ and C$_7$N$^-$}                                           
                                                                                   
  Table~\ref{tab:c5n} contains the results of the CAP-EOM-CCSD and EOM-CCSD calculations
  for the two longest cyanopolyyne anions considered in the present work:          
  C$_5$N$^-$ and C$_7$N$^-$.                                                       
  In C$_5$N$^-$, only the $^1\Pi$ state remains a resonance, with a very small width of less than 0.02 eV. Other valence
  excited states drop below the lowest detachment  threshold and become stable     
  with respect to autoionization.                                                  
  The recent theoretical                                                           
  study of the C$_5$N$^-$ photodetachment\cite{Douguet:CxNm:16} reported           
  the sharp $^1\Pi$ shape resonance at 5.1 eV, which agrees well with our calculations.
  Fortenberry in his EOM-CCSD calculations for C$_5$N$^-$ reported a valence bound $^1\Sigma^+$
  excited state at around 4.0 eV\cite{Fortenberry:JPCA:15}.                        
  Our results suggest that this is actually $^1\Delta$ excited state. Our symmetry assignment
  is consistent with an extremely  small oscillator strength ($\sim 10^{-6}$) for the
  transition to the anion ground state reported in Ref. \citenum{Fortenberry:JPCA:15}.
                                                                                   
  In C$_7$N$^-$ the lowest-lying                                                   
  $\pi \to \pi^\ast$ and $\sigma \to \pi^\ast$ transitions  are stabilized even  
  further, such that                                                               
  even the $^1\Pi$  state becomes bound.                                           
  Thus,  C$_7$N$^-$                                                                
  anion supports seven bound excited states, which is quite remarkable             
  for an anion. What is not shown in Table~\ref{tab:c5n} and Fig.~\ref{fig:endiag} 
  is a new low-lying resonance of $^1\Sigma^+$ symmetry emerging in C$_7$N$^-$.    
  This new state of a mixed valence-diffuse character is discussed            
  below, in Sec. \ref{sec:DBS} devoted to dipole-stabilized states.                
  We note that C$_7$N$^-$                                                          
  has additional low-lying valence resonances approaching the electron detachment threshold
  corresponding to excitations from deeper $\pi$ occupied                          
  orbitals. They are located about $~2$ eV above the lowest valence states of the same
  symmetry, which are reported                                                     
  in Table~\ref{tab:c5n}.                                                          
  Here we do not discuss these states, since our focus is on the lowest            
  transitions, i.e., originating  from the highest occupied $\pi$ or $\sigma$ orbitals.
                                                                                   
  \begin{table}[h!]                                                                
  \renewcommand{\arraystretch}{1.2}                                                
  \caption{Excitation energies (eV) of low-lying valence states of C$_5$N$^-$      
    and C$_7$N$^-$. Basis set is aug-cc-pVTZ+3s3p.                                 
    \label{tab:c5n}}                                                               
  \begin{tabular*}{0.48\textwidth}{@{\extracolsep{\fill}}lll}
  \hline                                                                           
  State &  C$_5$N$^-$  &  C$_7$N$^-$  \\                                           
  \hline                                                                           
  $^3\Sigma^+  (\pi \to \pi^\ast)$     &  3.03    &  2.48      \\                 
  $^3\Pi       (\sigma \to \pi^\ast)$  &  4.26    &  3.94      \\                 
  $^3\Delta    (\pi \to \pi^\ast)$     &  3.52    &  2.90      \\                 
  $^3\Sigma^-  (\pi \to \pi^\ast)$     &  3.85    &  3.17      \\                 
  $^1\Pi       (\sigma \to \pi^\ast)$  &  5.12$^{\textrm a}$    &  4.69      \\    
  $^1\Delta    (\pi \to \pi^\ast)$     &  3.98    &  3.26      \\                 
  $^1\Sigma^-  (\pi \to \pi^\ast)$     &  3.88    &  3.19      \\                 
  \hline                                                                           
  \end{tabular*}
    {\small $^{\textrm a}$This state is a resonance with the following parameters: $E_R^{(0)}=5.12\;$eV, $\Gamma^{(0)}=0.017\;$eV, 
  $v^{(0)}=1.1\times 10^{-4}$, $E_R^{(1)}=5.13\;$eV, $\Gamma^{(1)}=0.016\;$eV, $\eta_{\textrm{opt}}^{(1)}=2.4\times10^{-3}$,
    $v^{(1)}=1.9\times 10^{-5}.$}                                                   
  \end{table}
  \subsection{General trends in electronic spectra}                                
  \protect\label{sec:GeneralTrends}                                                
      \begin{figure*}[h!]
    \centering
    \includegraphics[width=5cm]{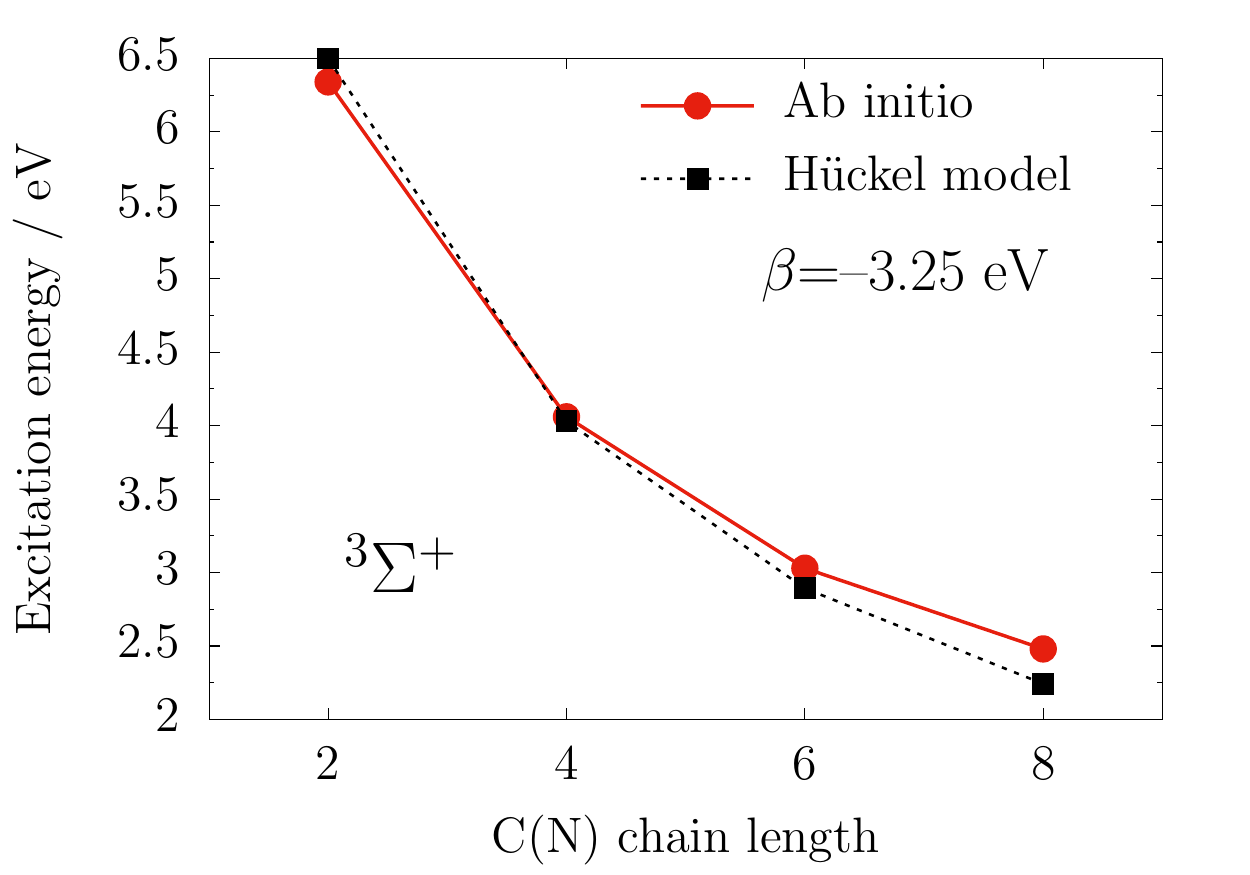}                                   
    \includegraphics[width=5cm]{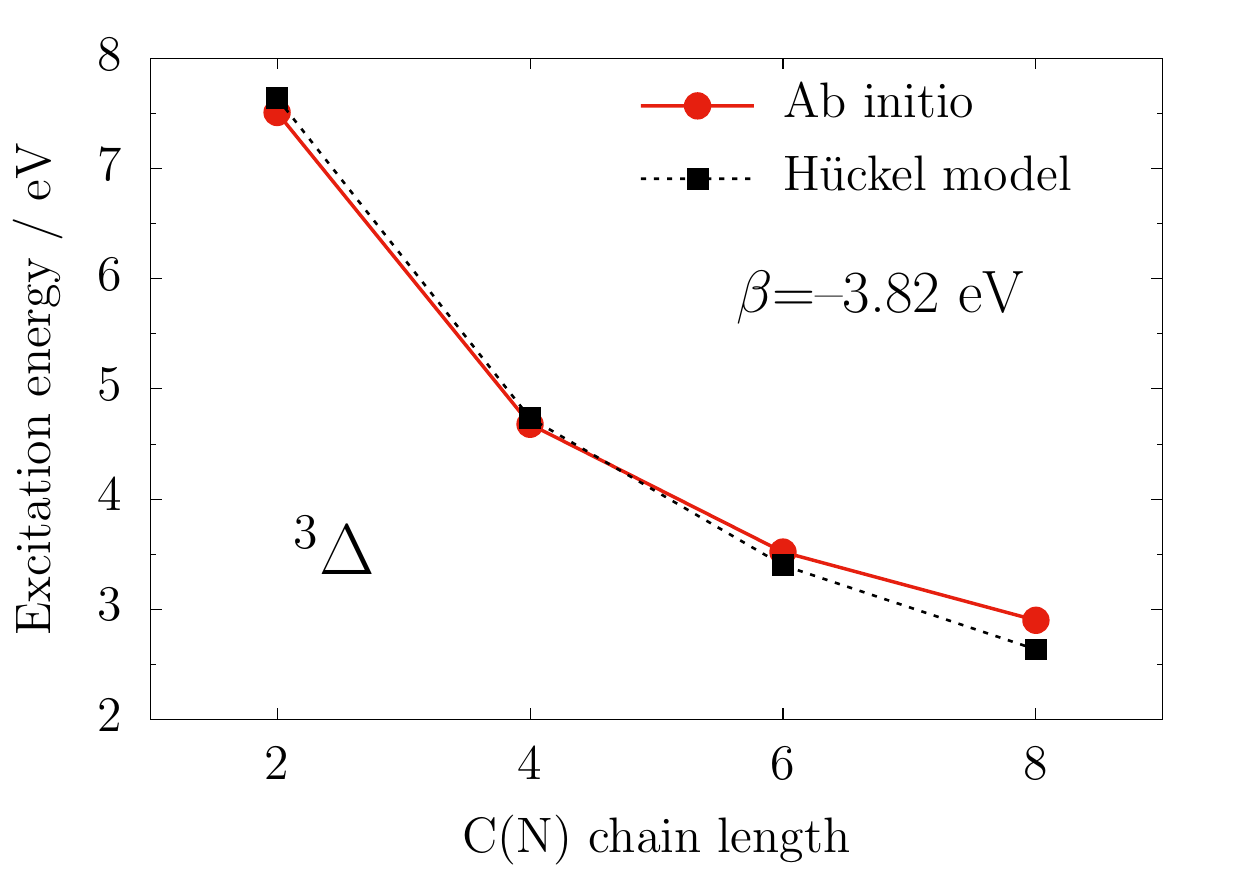}                                   
    \includegraphics[width=5cm]{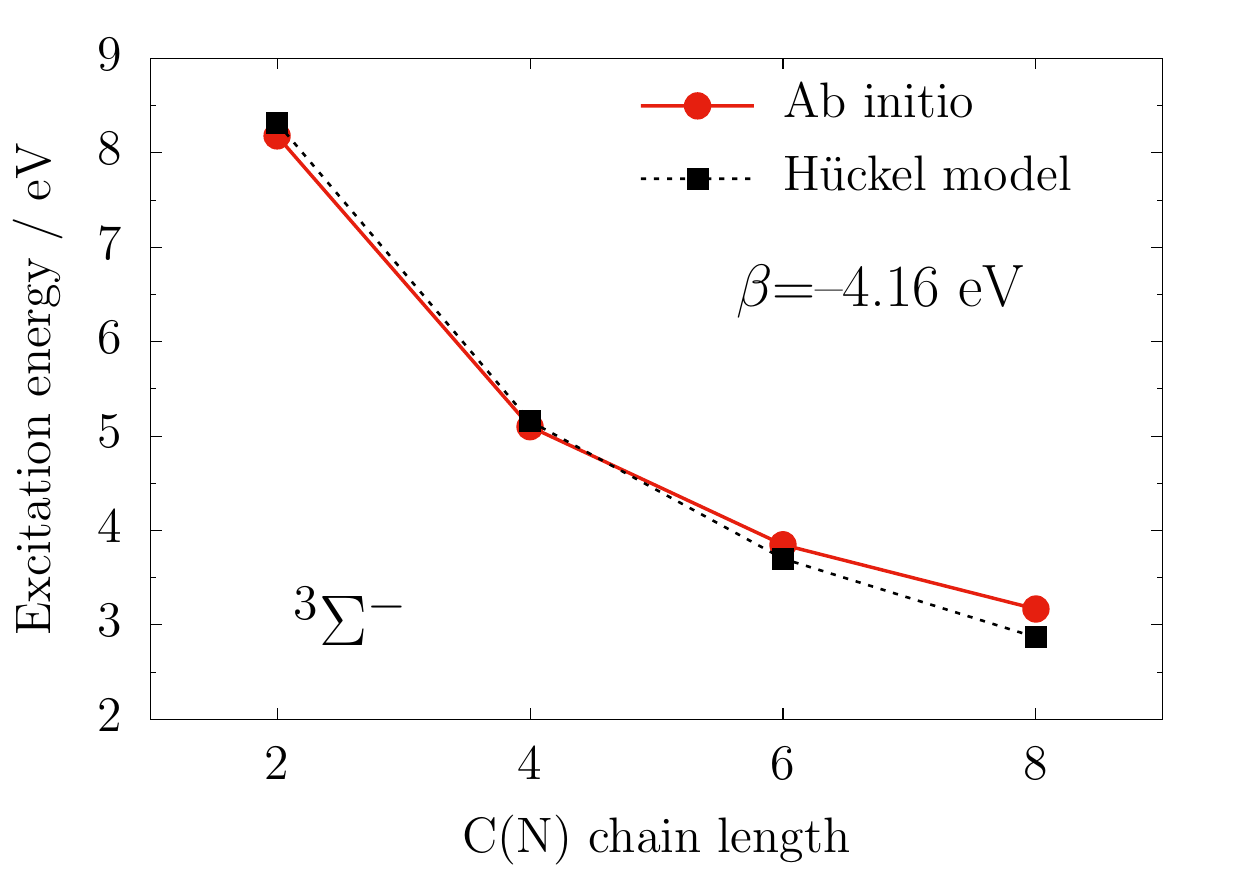}\\                                 
    \includegraphics[width=5cm]{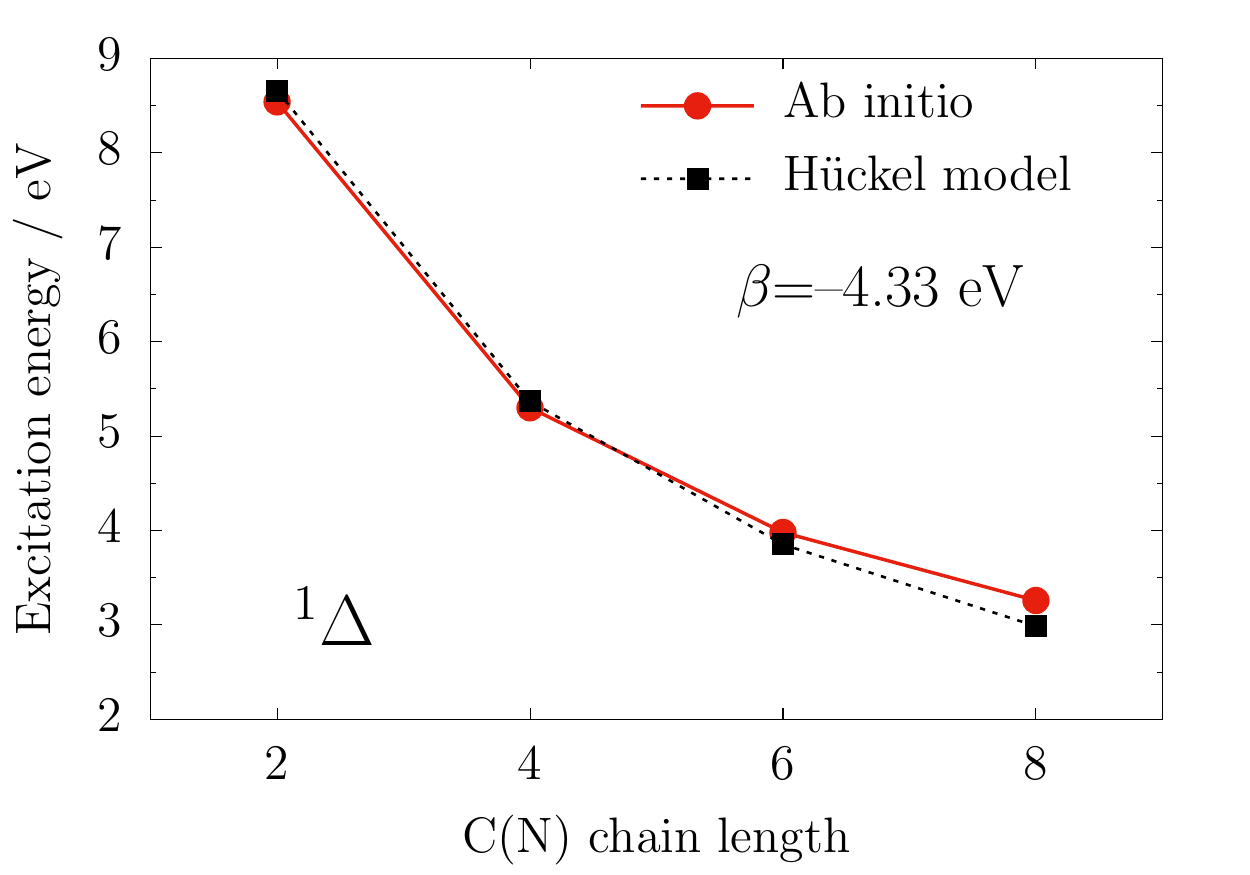}                                   
    \includegraphics[width=5cm]{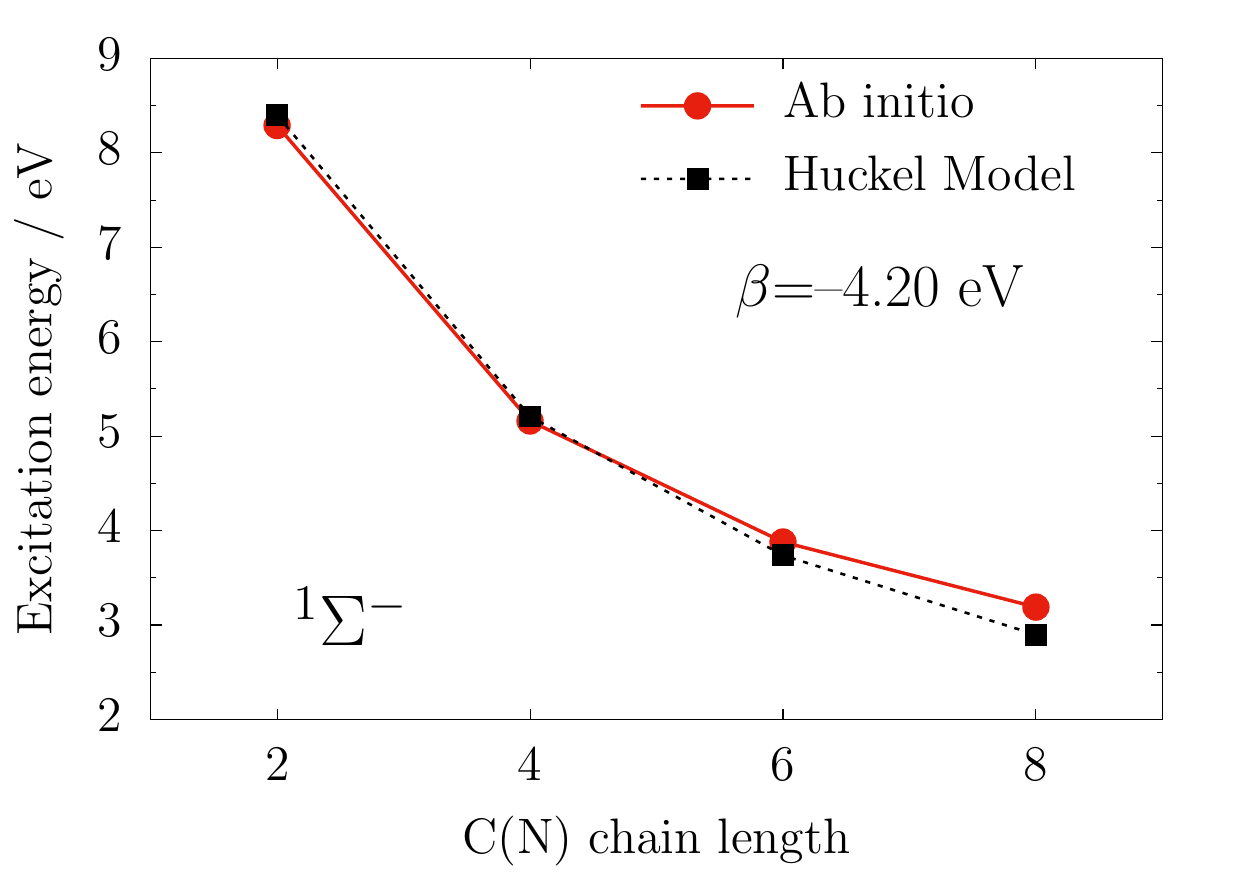}                                   
  \caption{                                                                        
    Excitation energies of $\pi\pi^\ast$ states                                   
    from {\it ab initio} calculations and from                                     
    the H\"{u}ckel model.                                                              
    \label{fig:huckel}}                                                            
  \end{figure*}                                                                                   
  It is instructive to  assemble  and compare the results                          
  for all  cyanopolyyne anions studied in the present work.                        
  Figure~\ref{fig:endiag} shows that the energies of all  $\pi \to \pi^\ast$ transitions
  decrease to a similar extent as the carbon chain lengths increases.              
  As mentioned above, this can be  rationalized in terms of the                    
  H\"{u}ckel model description of $\pi-$conjugated systems. According to the H\"{u}ckel's model,
  in a $\pi$ system comprised of $N$ atoms in a linear chain,                      
  the lowest $\pi \to \pi^\ast$ (HOMO-LUMO) transition energy reads:              
  \begin{equation*}                                                                
  e_{\rm{LUMO}}-e_{\rm{HOMO}}=-4\beta \sin{\frac{\pi}{2(1+N)}},                    
  \end{equation*}                                                                  
  and it depends only on one parameter $\beta$.                                    
  We  determined optimal $\beta$ values for each of the  $\pi \to \pi^\ast$ excitations
  by least squares fitting.                                                        
  Figure~\ref{fig:huckel}  compares  {\it ab initio} transition                    
  energies with  those from  the H{\"u}ckel model based on  fitted $\beta$'s       
  for each of the transitions.                                                     
  The optimal $\beta$ values                                                       
  vary between $-3.25$ and $-4.33$ eV.                                             
  The variation in $\beta$                                                         
  values is not surprising, as we describe multiple states of                      
  different spin and spatial symmetry with the same model.                         
  Similarly, different $\beta'$s are used to describe singlet/triplet transitions in
  conjugated hydrocarbon molecules\cite{Zahradnik:CCCC:71}.                        
  In all cases shown in Fig.~\ref{fig:huckel} there is a reasonably good agreement, highlighting
  the regularity in the                                                            
  excitation energies in the  cyanopolyyne anions.                                 
  Remarkably, the  H{\"u}ckel model  is capable of reproducing consistently        
  the transitions to both metastable and bound states.                             
  The agreement between the two curves in Fig.~\ref{fig:huckel}                    
  can be made almost perfect                                                       
  if one allows for slightly  different values of the resonance integrals $\beta$  
  for CN and CC bonds.

  The energies of metastable states are lowered                                    
  and their widths become more and more narrow                                     
  with the increasing chain length.                                                
  This behavior is explained by increased delocalization, which is                 
  captured by the Huckel model. In addition,                                       
  polarization, which increases in longer carbon chains,                           
  may also contribute towards resonance stabilization\cite{Bardsley:67,Krauss:70}.  Consequently, metastable states undergo gradual
  stabilization in longer species and eventually become bound.                     
  This trend is also illustrated in Fig.~\ref{fig:endiag} when the excited states shift down and
  drop below                                                                            
  the lowest neutral threshold (marked with black solid or dashed lines in         
  Fig.~\ref{fig:endiag}).                                                          
  As mentioned above, all CN$^-$ resonances are of shape type                      
  because they can autoionize via a one-electron transition. This is true for      
  all resonances located above both ($^2\Sigma$ and $^2\Pi$) neutral thresholds.   
  If a metastable state drops below its  parent neutral but is still  above the    
  other neutral threshold, then it becomes a  Feshbach                             
  resonance. We observe such situation only for the $^3\Delta$ state:              
  it is  a shape resonance in CN$^-$, Feshbach resonance in C$_3$N$^-$, and,       
  finally, a stable bound state in C$_5$N$^-$ and C$_7$N$^-$. All other metastable states
  are converted directly from shape resonances into stable bound states.           
                                                                                   
  Comparing widths of the singlet and triplet resonances of the same symmetry, we  
  note that, in all cases, the singlet states are broader than triplets.           
  A possible reason is that the singlet continuum  has a higher                    
  density of states than the triplet continuum due to Pauli exclusion principle    
  excluding some of the electronic configurations of the same-spin electrons.      
  Consequently, the coupling of                                                    
  singlet metastable states with the continuum is stronger and their lifetimes
  are shorter.
  
\begin{figure}[h!]
      \includegraphics[width=8cm]{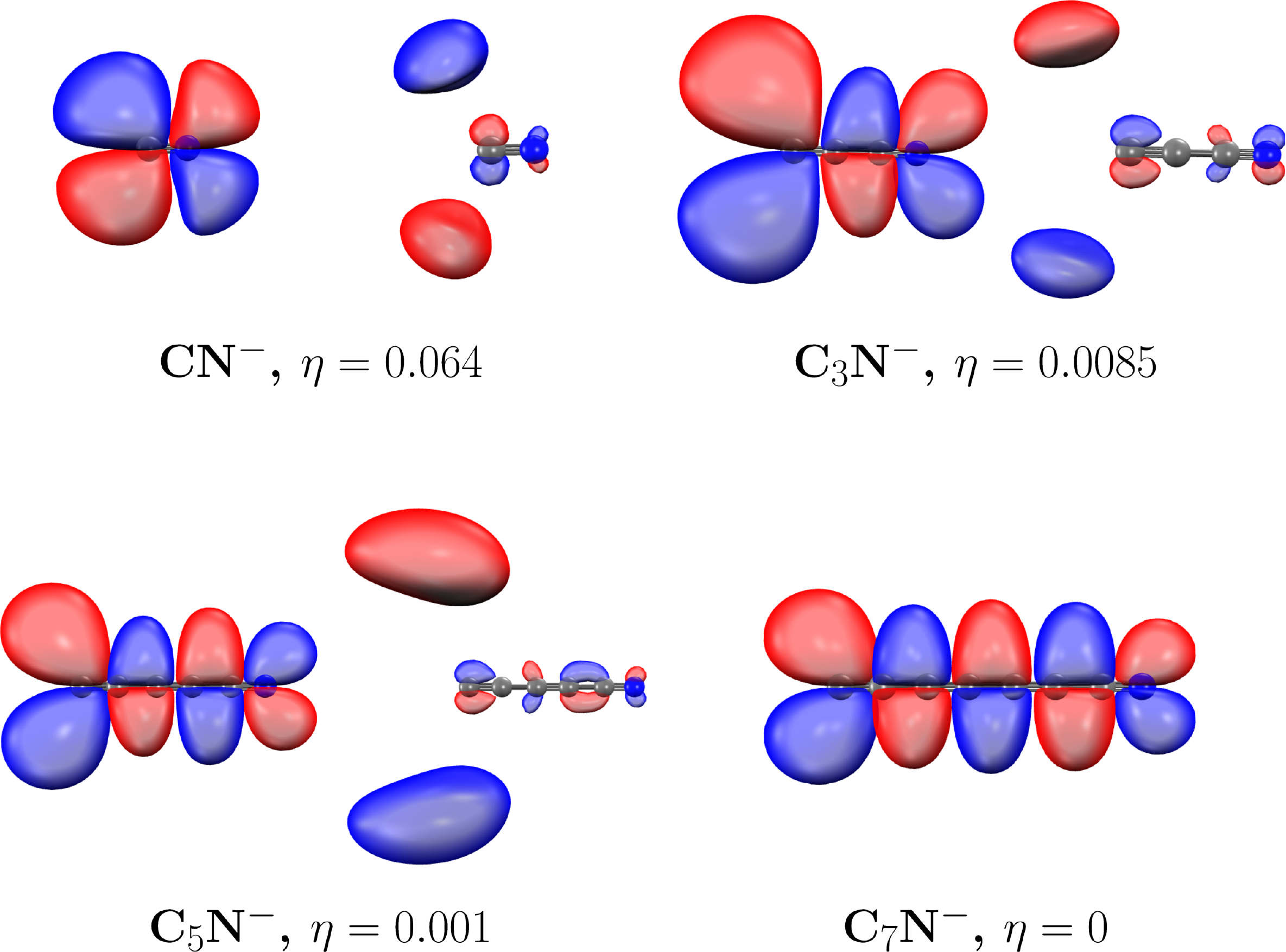}                                     
  \caption{                                                                        
    Real (left-hand side in each panel) and imaginary (right-hand side in each panel) parts of Dyson orbitals for the $^1\Pi$ state in
    cyanopolyyne anions. Dyson orbitals were calculated as the overlap with the lowest       
    $^2\Sigma^+$ state of the corresponding parent radical.                        
    \label{fig:dyscom}}                                                                          
  \end{figure}

  A common feature of all valence resonances and bound states shown in             
  Fig.~\ref{fig:endiag}                                                            
  is that they are dominated by one-electron excitation into the lowest            
  unoccupied $\pi^\ast$                                                           
  orbital. This is best illustrated by the Dyson orbitals of the                   
  excited states calculated as overlaps with with the corresponding parent         
  radical ($^2\Pi$ neutral for $^{3}\Sigma^+$, $^{1/3}\Sigma^-$ and $^{1/3}\Delta$, or
  $^2\Sigma^+$ neutral for  $^{1/3}\Pi$ excited states). Figure~\ref{fig:dyscom} shows
  illustrative                                                                     
  Dyson orbitals for the identified $^1\Pi$ states (either metastable or bound).   
  Both real and imaginary parts of all   Dyson orbitals                            
  have typical nodal structure of $\pi^\ast$ orbitals. Their norms\cite{DOnorms} are almost purely
  real and equal to $\sim 0.47-0.002i$  ($\Pi$  states) and $0.24+0.003i$          
  ($\Sigma^+$, $\Sigma^-$ and $\Delta$ states).                                    
  Relatively small norms of the                                                    
  Dyson  orbitals indicate importance of correlation.
  
   \begin{table*}[b!]
  \renewcommand{\arraystretch}{1.2}                                                
  \caption{Dipole-stabilized and diffuse states of $\Sigma^+$ symmetry in cyanopolyyne anions.
  Metastable states (in C$_5$N$^-$, C$_7$N$^-$) are characterized by               
  position $E_R$, width $\Gamma$ (in eV), optimal value                            
  of $\eta$ parameter, and corresponding trajectory velocity (in atomic            
  units). For DBS (C$_3$N$^-$) we report binding energy.                           
  Basis sets are: aug-cc-pVTZ+8s5p1d (C$_3$N$^-$)                                  
  and aug-cc-pVDZ+3s3p (C$_5$N$^-$ and C$_7$N$^-$).                                
  Numbers in parentheses denote powers of 10.                                      
  \label{tab:diffuse}}
  \begin{tabular*}{\textwidth}{@{\extracolsep{\fill}}llllllllll}
  \hline                                                                           
  Anion    &  \multicolumn{1}{c}{State}  &  $E_R^{(0)}$  &  $\Gamma^{(0)}$  &   $\eta_{\textrm{opt}}^{(0)}$  & $v^{(0)}$
                     &  $E_R^{(1)}$  &  $\Gamma^{(1)}$  &   $\eta_{\textrm{opt}}^{(1)}$  & $v^{(1)}$  \\
  \hline                                                                           
  C$_3$N$^-$  & $2^3\Sigma^+(\sigma \to \sigma^\ast)$    & --0.006$^{\textrm a}$
  &  ---  &            &          &          &         &         &         \\      
           &    $2^1\Sigma^+(\sigma \to \sigma^\ast)$    & --0.002$^{\textrm a}$    &  ---  &            &          &          &         &         &         \\
                                                                                   
  C$_5$N$^-$   & $2^3\Sigma^+(\sigma \to \sigma^\ast)$    &  5.63     &  0.78 &    0.02    & 6.6(-3)  &  5.54    &  0.53   &  0.04   & 9.5(-4) \\
               & $2^1\Sigma^+(\sigma \to \sigma^\ast)$    &  5.76     &  0.76 &    0.018   & 6.4(-3)  &  5.64    &  0.56   &  0.034  & 2.3(-3) \\
  C$_7$N$^-$   & $2^3\Sigma^+(\sigma \to \sigma^\ast)$    &  5.16     &  0.34 &    0.001   & 3.5(-3)  &  5.12    &  0.26   &  0.002  & 1.3(-3) \\
               & $2^1\Sigma^+(\pi \to \pi^\ast)$          &  5.33     &  0.13 &    0.03    & 1.3(-3)  &  5.30    &  0.14   &  0.03   & 1.4(-3) \\
  \hline                                                                           
  \end{tabular*}
  {\small $^{\textrm a}$Binding energy (in eV) with respect to the lowest ionization threshold.}
  \end{table*}
  
  Of all the identified valence resonances, the $^1\Pi$  state                         
  is the most interesting, as it                                                   
  is  directly accessible via one-photon transition from the $^1\Sigma^+$ ground   
  state of the anion.                                                              
  This state can be observed in one-photon                                         
  photodetachment (or absorption cross) section.                                   
  Also, it can serve as a gateway in a two-step radiative electron attachment,     
  a process discussed in the context of possible mechanisms of                     
  C$_{2n-1}$N$^-$ formation in the ISM.
  
    \subsection{In the search of dipole-stabilized states in cyanopolyynes}          
  \protect\label{sec:DBS}

  The discussion above focused exclusively on valence excited                      
  states. Let us now discuss states derived by                                     
  excitation into diffuse orbitals and stabilized by the                           
  electrostatic interaction between the outer electron and the dipole moment of the neutral core. These can be
  either dipole-bound states or dipole-stabilized resonances.                      
  According to the  results in Table~\ref{tab:vde}, C$_3$N$^-$ is most likely to   
  support dipole bound states                                                      
  because its parent radical in the electronic ground state has dipole moment      
  exceeding the critical value\cite{Crawford:DBS:70} of $\sim$2.5 D.               
  In longer anions, DBS would require excitation                                   
  from a lower-lying $\sigma$ orbital.                                             
  To locate  DBS, we                                                               
  carried out EOM-EE-CCSD calculations\cite{DBS}                                   
  with extended basis sets which included                                          
  up to 8$s$, 5$p$ and 1$d$ additional diffuse functions on                        
  top of the aug-cc-pVTZ basis set.
    \begin{figure*}[h!]                                                               
  \includegraphics[scale=0.60]{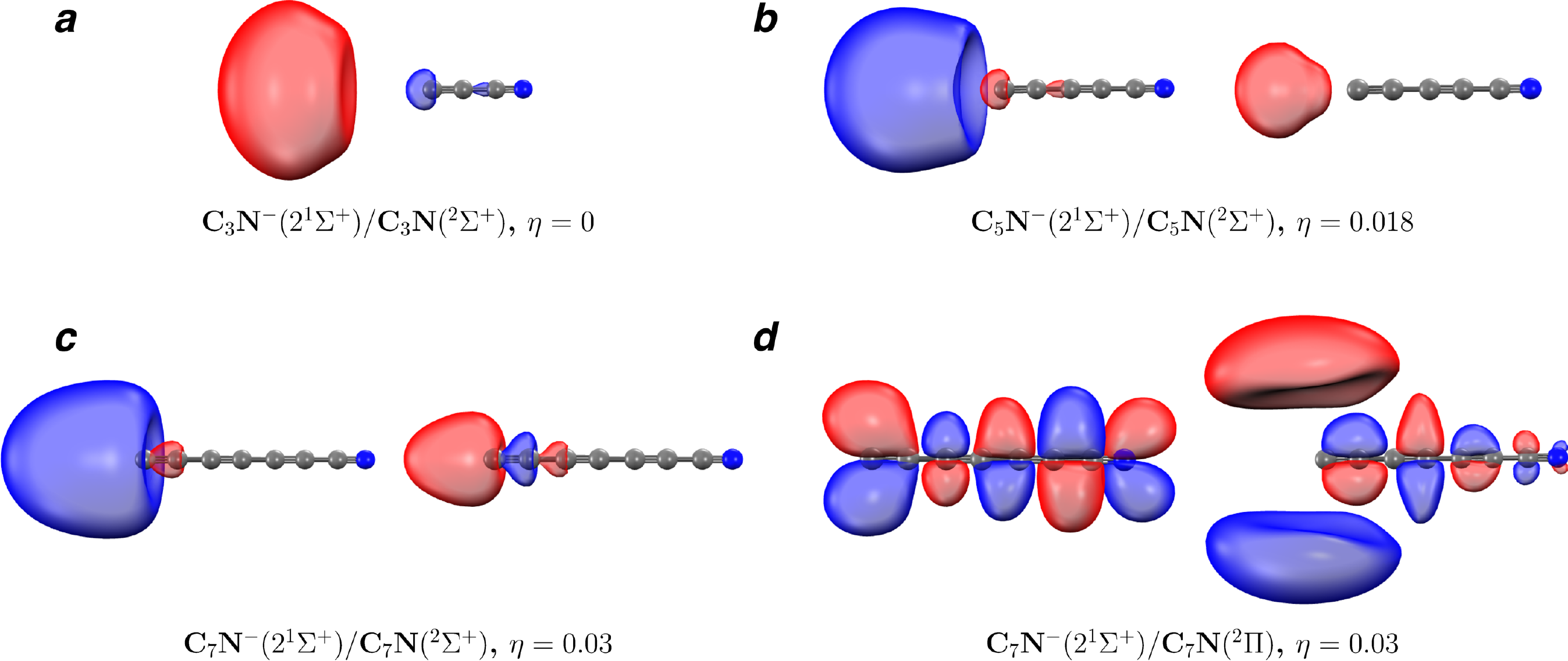}                                     
  \caption{                                                                        
    Real (left-hand side in each panel) and imaginary (right-hand side in each panel) parts of Dyson orbitals for diffuse $^1\Sigma^+$ states
    reported in Table~\ref{tab:diffuse}. Dyson orbitals were calculated as the     
    overlap with the $^2\Sigma^+$ state (panels {\it a--c}) or the $^2\Pi$ state (panel {\it d\;})
    of the corresponding parent radical.                                           
    \label{fig:dysdiff}}                                                           
  \end{figure*}
  In C$_3$N$^-$ we found two  weakly bound states ($^3\Sigma^+$ and $^1\Sigma^+$), 
  located just below the                                                           
  lowest electron detachment threshold.                                            
  Our best estimates of the vertical binding                                       
  energies (computed at the equilibrium structure of the anion)                    
  are 0.006 eV for the $^3\Sigma^+$ state and 0.002 eV for the                     
  $^1\Sigma^+$ state, obtained with the aug-cc-pVTZ+8s5p1d basis set.              
  Interestingly, although                                                          
  the parent radical C$_3$N has considerably large dipole moment ($\sim 3.9\;$D),  
  the binding energies of the DBS states are extremely small.                      
  For example, these binding energies are 2-6 times smaller than in DBS in         
  closely related HC$_3$N$^-$, which has closed-shell precursor                    
  with $\mu$=4.2 D.                                                                
  We attribute these weak stabilization energies to                                
  the open-shell character of the parent radical.                                  
  Detailed                                                                         
  discussion of this issue  will be presented elsewhere.                           
  In contrast to the previous  EOM-CCSD studies\cite{Fortenberry:JPCA:15,Crawford:DBS:11}
  in which the excitation energies of putative DBS states never dropped  below the detachment
  threshold, in our calculations  the identification of DBS as bound states        
  is unambiguous.                                                                  
  We did not find DBS in other cyanopolyynes (CN$^-$, C$_5$N$^-$ and C$_7$N$^-$),  
  which is consistent with low values of dipole moments of the respective neutral precursors
  in their electronic ground states.                                               
  However, the CAP-augmented EOM-CCSD calculations for C$_5$N$^-$ and C$_7$N$^-$   
  identified  low-lying excited states of $\Sigma^+$ symmetry, which can be
  described as dipole-stabilized resonances.                                     
  These states, dominated                                                          
  by the excitation from frontier $\sigma$ into diffuse $\sigma^\ast$ orbital,    
  appear between 5.1 and 5.8 eV above the anionic ground                           
  state and have relatively large widths spanning  from $\sim0.3$ to $\sim 0.8$ eV 
  (see Table~\ref{tab:diffuse} for states denoted as  $2^1\Sigma^+$ and $2^3\Sigma^+$).
  The dipole-stabilized character of these states is revealed  by the shapes of    
  Dyson orbitals (calculated as the overlap with the $^2\Sigma^+$ state of parent  
  radicals). As shown in Fig.~\ref{fig:dysdiff}{\it a--c}, the excess electron resides
  outside the terminal carbon, i.e.,  at the positive end of the dipole.           
  The  $2^1\Sigma^+$ state of C$_7$N is an exception in this series,               
  as it  has a  mixed  $\pi \to \pi^\ast$ and $\sigma \to \sigma^\ast$           
  character. This mixing is reflected by relatively large c-norm values  of Dyson orbitals
  computed for the two different                                                   
  detachment channels corresponding to the                                         
  $^2\Pi$ (c-norm is $0.20-0.02i$,  Fig.~\ref{fig:dysdiff}$d$) and   $^2\Sigma^+$ (c-norm is $0.08+0.04i$, Fig.~\ref{fig:dysdiff}$c$) states of the radical.
  For other states, there is only one Dyson orbital with non-negligible c-norm.    
  Significant  $\pi \to \pi^\ast$ character of the $2^1\Sigma^+$ state in         
  C$_7$N suggests its classification as a valence state. Therefore, this resonance 
  is the first appearance of $^1\Sigma^+$ excitation from the $\pi \to \pi^\ast$  
  manifold,  which was missing in shorter cyanopolyyne chains (see                 
  Fig.~\ref{fig:endiag}).

  The shapes of $\eta$-trajectories for the resonances of $\sigma \to \sigma^\ast$
  character differ considerably                                                    
  from those for valence-type states ($\eta$-trajectories for all metastable states are given in Supplementary Information).
  The $\eta$-trajectories for $\Sigma^+$ dipole-stabilized resonances are more difficult
  to interpret as they feature quasi-oscillatory behavior                          
  with several stabilization points. Optimal $\eta_{\rm opt}$ values               
  reported in Table~\ref{tab:diffuse} correspond to the lowest $\eta$ at which stabilization is
  manifested both by the minimum of the trajectory velocity, Eq.~(\ref{eq:vel}), and by
  constant size of the wave function, defined as $\langle R^2 \rangle$.            
                                                                                   
  Because the dipole moment of the CN radical is small (1.35 D), it is unlikely that this molecule would support
  dipole-stabilized anionic states. Quite surprisingly, the CAP-augmented EOM-CCSD 
  calculations yield some low-lying diffuse                                         
  metastable states (around 5.7 eV), derived by the excitation from $\sigma$ orbital into diffuse
  $\sigma$-like orbitals.
  These states persists in CAP-EOM-CCSD calculations with larger basis sets and different CAP onsets.   
  Similar states appear in some CBF calculations (e.g., with aug-cc-pVTZ+3s3p basis set), 
  but they all disappear when larger basis set (aug-cc-pVTZ+6s6p
  or aug-cc-pVTZ+9s9p) is employed in CBF EOM-CCSD. 
  We concluded that these states are false resonances
  appearing as artifacts of the                                                               
  theoretical approaches.
  
  \subsection{Possible experimental observation of the predicted features}         
                                                                                   
  \begin{figure*}[b!]                                                               
  \centering
    \includegraphics[width=5.5cm]{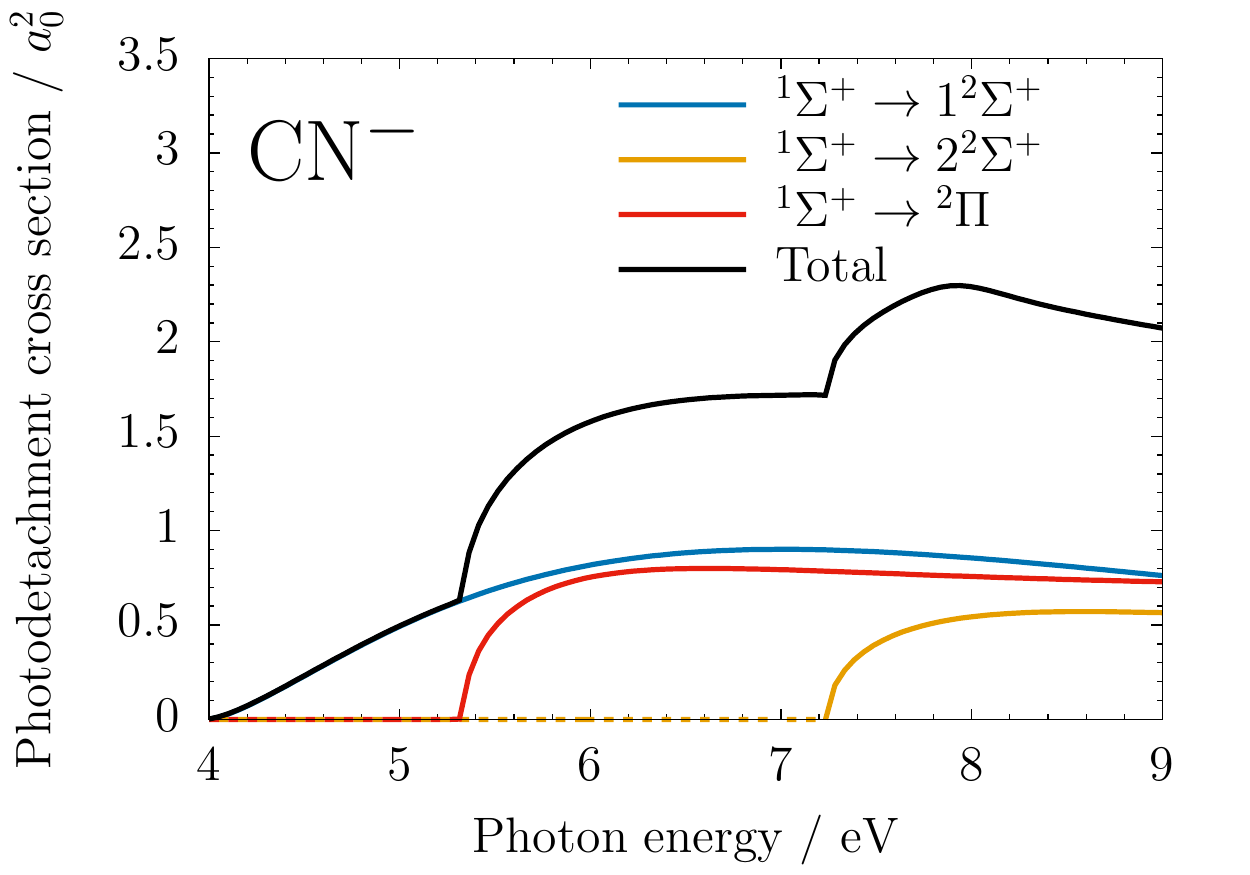}\hspace{0.5cm}                                   
    \includegraphics[width=5.5cm]{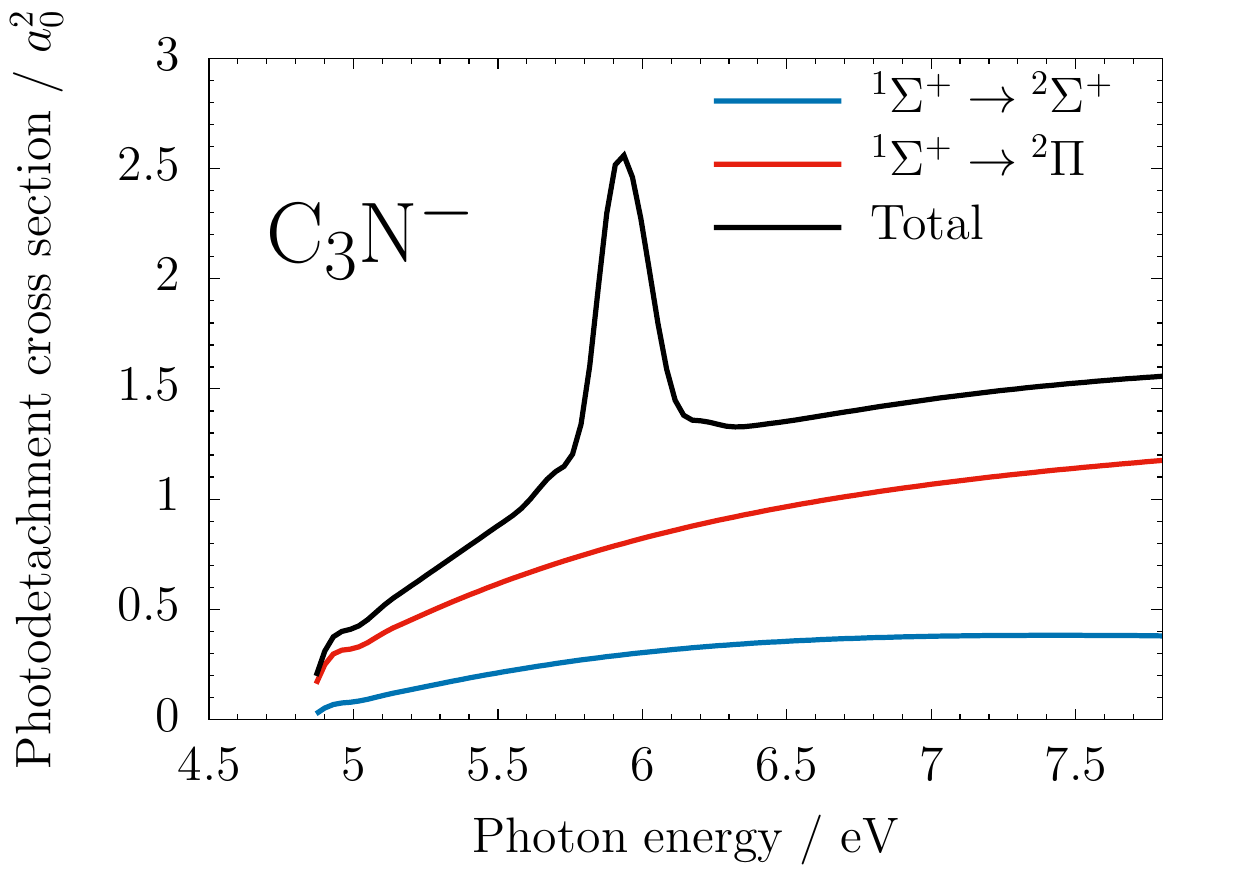}\hspace{0.5cm}                                   
    \includegraphics[width=5.5cm]{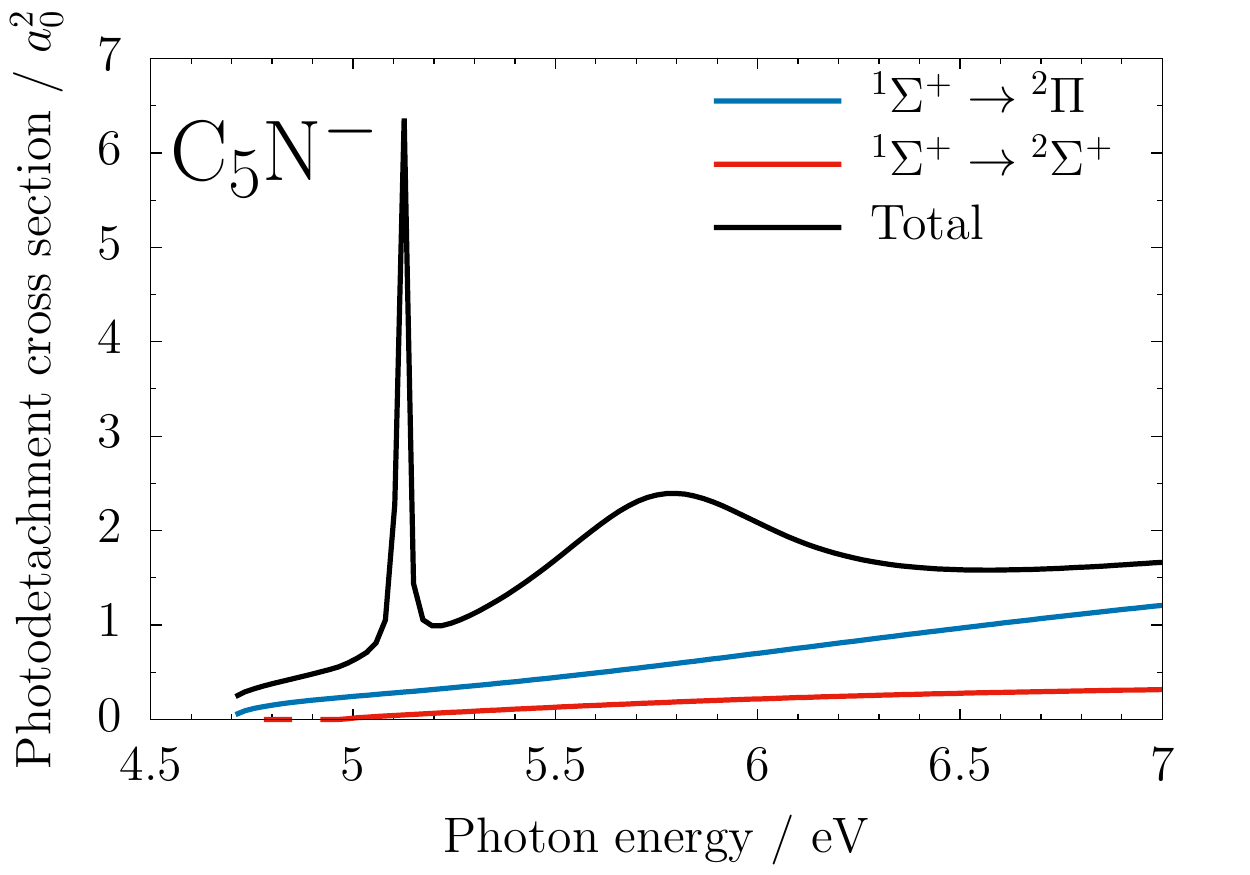}                                     
  \caption{                                                                        
    Calculated photodetachment cross sections for CN$^-$, C$_3$N$^-$,  and         
    C$_5$N$^-$. Contribution from detachment to $^2\Sigma^+$                       
    and $^2\Pi$ neutral states are shown separately. The sharp peaks in the total cross section
    are due to the $^1\Pi$ resonance.                                              
  \label{fig:cs}}                                                                  
  \end{figure*}                                                                     
                                                                                   
  There have been yet no experimental observation of the electronic resonances  in 
  cyanopolyyne anions. Below we discuss how our                                    
  predictions regarding electronic structure of C$_n$N$^-$ can be verified         
  experimentally. The easiest to observe would be  the                             
  low-lying  $^1\Pi$ and  $^1\Sigma^+$ resonances,                                 
  as they are  dipole-coupled with the anionic $^1\Sigma^+$ ground state and       
  should, therefore, be visible in photodetachment cross section.                  
  The calculated photodetachment  cross sections are                               
  shown in Fig.~\ref{fig:cs},                                                      
  including the contribution due  to the $^1\Pi$ and  $^1\Sigma^+$                 
  resonances. Details of photodetachment  cross section                            
  computations are given in the Supplementary Information.                         
  The predicted resonance peaks due to the $^1\Pi$ resonance in C$_3$N$^-$ and C$_5$N$^-$ are rather prominent
  and should be easily detected if the photodetachment cross section               
  is measured in the appropriate energy range.                                     
  Similar resonance features in the photodetachment cross                          
  section of C$_3$N$^-$ and C$_5$N$^-$ have been predicted in Ref. \citenum{Douguet:CxNm:16}.
  For CN$^-$  the large widths of the $^1\Pi$  resonance makes its                 
  contribution to the total photodetachment cross section less pronounced and      
  most of the features  in the total cross section come from opening new detachment
  channels.                                                                        
                                                                                   
  Whereas the $^1\Pi$ and $^1\Sigma^+$ resonances could be accessed  directly from the  electronic ground
  state of the C$_{2n-1}$N$^-$ anions,  other resonances,                          
  both singlets and triplets,  could manifest themselves in                        
  electron scattering from the neutral C$_{2n-1}$N precursors, similarly to        
  the resonances                                                                   
  observed in electron impact experiments involving  open-shell molecules such as  
  NO\cite{Jelisavcic:PRL:03},                                                      
  (C$_2^-$)\cite{Pedersen:PRL:98}, or C$_2$F$_5$\cite{Haughey:JCP:12}              
  (via dissociative electron attachment process).                                  
  One can expect prominent features in the total                                       
  cross section for shape resonances associated with electron                      
  being captured by the  C$_{2n-1}$N ground state ($^1\Pi$ and $^3\Pi$ resonances in CN$^-$ or
  C$_3$N$^-$).                                                                     
  Less pronounced peaks would  appear for                                          
  other shape resonances originating from the excited state of C$_{2n-1}$N radical 
  ($^3\Sigma^+$,  $^1\Delta$, $^3\Delta$, $^1\Sigma^-$, or  $^3\Sigma^-$ states in CN$^-$).
  In this case, resonances would preferably decay to its parent                    
  radical  in the excited state ($^2\Pi$ in CN$^-$), thus enhancing the probability
  of the formation of  excited state  C$_{2n-1}$N (Ref. \citenum{Allan:HCA:82}).
  
  Bound excited states of cyanopolyynes can be                                     
  observed by absorption or emission spectroscopy. The first state of this kind    
  is $^3\Sigma^+$ in C$_3$N$^-$. While the optical                                 
  transition from the $X^1\Sigma^+$ ground state is spin-forbidden,                
  this state is most likely responsible for the observed  long-lived               
  phosphorescence in C$_3$N$^-$, with the band origin at 3.58 eV in rare gas       
  matrix\cite{Turowski:08} (our vertical                                           
  excitation energy is 4.06 eV). Similar emission bands assigned to                
  the $X^1\Sigma^+ \leftarrow {^3\Sigma^+}$ transition  were measured for          
  the neutral HC$_5$N and HC$_7$N molecules\cite{Turowski:10,Couturier:14}         
  with origins at 2.92 and 2.45 eV, respectively.                                  
  These values compare favorably with our predictions for the $^3\Sigma^+$         
  state in the isoelectronic  C$_5$N$^-$  and C$_7$N$^-$ anions (vertical          
  excitation energies of 3.03 and 2.48 eV).                                        
  Bound excited states of $^1\Sigma^+$ and $^1\Pi$ symmetry can be directly        
  probed by absorption spectroscopy.                                               
  The absorption spectrum of C$_7$N$^-$ embedded in  neon matrices\cite{Grutter:JCP:99} reveals
  electronic transition with the origin at 4.53 eV.                                
  The authors assigned this transition as                                          
  $2^1\Sigma^+ \leftarrow                                                          
  X^1\Sigma^+$,  by analogy with HC$_{2n}$H.                                       
  Our results suggest that this                                                    
  feature more likely corresponds  to the $^1\Pi \leftarrow                        
  X^1\Sigma^+$ transition  (vertical excitation energy of  4.69 eV, according to our calculations).
  However, we also predict a  narrow $2^1\Sigma^+$ resonance at $\sim 5.1$ eV and  
  lifetime of 5 fs,                                                                
  with large transition dipole moment to the anionic ground state (see Table SIII  
  in the Supplementary  Information). Therefore,                                   
  the original assignment of the observed band as $2^1\Sigma^+ \leftarrow   X^1\Sigma^+$ cannot be
  ruled out, since neon matrix can stabilize                                       
  the excited states and convert the resonance into a bound state.                 
  Bound excited states in anionic species can also be probed by                    
  multiphoton resonance-enhanced  detachment spectroscopy, as it was               
  demonstrated for polyatomic carbon anions\cite{Zhao:96}.                         
  In this technique bound excited states serve as intermediate states in           
  multiphoton detachment.                                                          
  In principle, absorption spectroscopy                                            
  can also detect electronic resonances if their lifetime is sufficient\cite{Tulej:00}.
  For example, the sharp $^1\Pi$ resonance in C$_5$N$^-$ with detachment           
  lifetime of 40 fs can lead to discrete absorption band, despite being embedded   
  in the continuum.
  
  \section{Summary and conclusions}                                                
  Recent discovery of several carbon chain anions in the ISM triggered             
  theoretical and experimental studies  aiming to  understand their                
  chemical and spectroscopic properties. Such knowledge is essential               
  to explain the molecular origin of these species in space and                    
  might also guide astrochemical search for new, yet undiscovered  anions.         
  In this contribution, we presented a systematic {\it ab initio} study of the     
  electronic structure of                                                          
  cyanopolyyne anions C$_{2n+1}$N$^-$ ($n=0,\dots,3$), with the focus on           
  their low-lying metastable (with respect to electron detachment) and bound excited states.
  The calculations were carried out  by the CAP-EOM-CCSD method, which             
  describes bound and metastable states on an equal footing.                             
  Already for CN$^-$, the smallest anion in the series,  we found several          
  low-lying singlet and triplet metastable states ($^1\Sigma^-$, $^1\Delta$,       
  $^1\Pi$, $^3\Sigma^+$, $^3\Sigma^-$, $^3\Delta$, $^3\Pi$). These                 
  metastable states can be classified as shape resonances                          
  and correspond to valence $\pi\to\pi^\ast$ or $\sigma\to\pi^\ast$ excitation   
  form the ground state of the anion. Dominant one-electron character              
  of these states  is clearly manifested in the shape of the respective Dyson orbital, 
  which in all cases resembles a typical  $\pi^\ast$ orbital.                     
  In longer species, the positions of all valence metastable states are lowered    
  and their widths decrease; eventually, these resonances become bound.            
  The stabilization  can be explained by                                           
  the increased delocalization of the frontier molecular orbitals.
   On the route from  metastable to stable bound states,                            
  some resonances might change their character from                             
  shape type to Feshbach type, as it                                               
  happens for the $^3\Delta$ state in C$_3$N$^-$ anion.                            
  For all valence resonances, the                                                  
  singlets are systematically broader than the corresponding triplets              
  of the same symmetry, indicating stronger coupling  between                      
  the bound and  continuum parts of the spectrum for singlet spin symmetry.        
                                                                                   
  In the case of excited states of  $\pi \to \pi^\ast$                            
  character (i.e. $\Sigma^+$, $\Sigma^-$ and  $\Delta$) trends in  excitation      
  energy as a function of carbon chain length are well reproduced  by the  H\"{u}ckel model.
  Thus, the H\"{u}ckel model retains its validity across the domains of bound and metastable states.
                                                                                   
  Apart from the valence metastable and bound excited states,                       
  some of cyanopolyyne anions can also support dipole-stabilized states, either    
  bound or metastable. Their existence                                             
  depends on the magnitude of dipole moment of the parent neutral                  
  radical. Our EOM-EE-CCSD calculations showed unequivocally                       
  that C$_3$N$^-$ has near-threshold bound states  of $^1\Sigma^+$                 
  and $^3\Sigma^+$ symmetry, bound vertically by less than 0.01 eV.                
  In longer  cyanopolyyne anions (C$_5$N$^-$ and C$_7$N$^-$) the CAP calculations identified
  relatively broad dipole-stabilized resonances.
  
  The resonances identified in this work are located in the range from a  few tenths
  of  eV (C$_5$N$^-$ and C$_7$N$^-$) up to                                         
  a few eV (CN$^-$) above the ground state of the parent neutral radical.          
  In  C$_5$N$^-$ the                                                               
  $^1\Pi$ resonance appears in the energy range that might be relevant for its  formation
  via REA, especially if we                                                        
  take into account the effect of structural relaxation.                           
  In the case of C$_3$N$^-$ 
  rovibrational resonances associated with the $^1\Sigma^+$ DBS provide a
  natural doorway for the formation of the anions in the ISM via electron attachment. 
  In this process the electron is  captured into DBS Feshbach vibrational level located
  near the detachment threshold,  and then the resulting transient  state
  can undergo stabilization to the anion's electronic ground state via
  dipole-allowed transition.
For the smallest interstellar anions, CN$^-$, the electronic structure calculations
  do not                                                                           
  support mechanisms of its formation involving direct electron                    
  capture by the CN radical.                                                       
  We concur with the previous studies\cite{Satta:AJ:15,Graupner:AJL:95,Graupner:NJP:06}
  that                                                                             
  chemical reactions and dissociation of                                           
  larger closed-shell molecules such as NC$_3$N or HC$_3$N                         
  are more likely sources of CN$^-$ occurrence in                                  
  space.                                                                           
                                                                                   
  We discussed several possible ways of how the predicted resonances can              
  manifest themselves in experimental observables, including                       
  photodetachment cross section, electron impact spectroscopy, and  absorption
  or emission spectra (in the case of stable bound states). The easiest to observe are
  the $^1\Pi$ and $^1\Sigma^+$ states, which have non-vanishing transition dipole moment  to the
  anion ground state; therefore, they can be probed by one-photon spectroscopy     
  of the anion ground state.
  
  \section*{Conflicts of interest}
  A.I.K. is a member of the Board of Directors and a part-owner of
  Q-Chem, Inc.
  
  \section*{Acknowledgments}
  This work has been supported in Los Angeles by the Army Research Office through grant
  W911NF-16-1-0232 and the Alexander von Humboldt Foundation (Bessel Award to A.I.K.).
  We thank Dr. Thomas Jagau                                                        
  and Prof. Ken Jordan for stimulating discussions and feedback about the manuscript.

\balance
\bibliography{all_ref1}

\providecommand*{\mcitethebibliography}{\thebibliography}
\csname @ifundefined\endcsname{endmcitethebibliography}
{\let\endmcitethebibliography\endthebibliography}{}
\begin{mcitethebibliography}{109}
\providecommand*{\natexlab}[1]{#1}
\providecommand*{\mciteSetBstSublistMode}[1]{}
\providecommand*{\mciteSetBstMaxWidthForm}[2]{}
\providecommand*{\mciteBstWouldAddEndPuncttrue}
  {\def\EndOfBibitem{\unskip.}}
\providecommand*{\mciteBstWouldAddEndPunctfalse}
  {\let\EndOfBibitem\relax}
\providecommand*{\mciteSetBstMidEndSepPunct}[3]{}
\providecommand*{\mciteSetBstSublistLabelBeginEnd}[3]{}
\providecommand*{\EndOfBibitem}{}
\mciteSetBstSublistMode{f}
\mciteSetBstMaxWidthForm{subitem}
{(\emph{\alph{mcitesubitemcount}})}
\mciteSetBstSublistLabelBeginEnd{\mcitemaxwidthsubitemform\space}
{\relax}{\relax}

\bibitem[Simons(2008)]{Simons:08:MolAnions}
J.~Simons, \emph{J. Phys. Chem. A}, 2008, \textbf{112}, 6401--6511\relax
\mciteBstWouldAddEndPuncttrue
\mciteSetBstMidEndSepPunct{\mcitedefaultmidpunct}
{\mcitedefaultendpunct}{\mcitedefaultseppunct}\relax
\EndOfBibitem
\bibitem[Simons(2011)]{Simons:ARPC:11}
J.~Simons, \emph{Annu. Rev. Phys. Chem.}, 2011, \textbf{62}, 107--128\relax
\mciteBstWouldAddEndPuncttrue
\mciteSetBstMidEndSepPunct{\mcitedefaultmidpunct}
{\mcitedefaultendpunct}{\mcitedefaultseppunct}\relax
\EndOfBibitem
\bibitem[Simons(2000)]{Simons:book:00}
J.~Simons, \emph{Photoionization and Photodetachment}, World Scientific
  Publishing Co., Singapore, 2000, vol. 10, Part II\relax
\mciteBstWouldAddEndPuncttrue
\mciteSetBstMidEndSepPunct{\mcitedefaultmidpunct}
{\mcitedefaultendpunct}{\mcitedefaultseppunct}\relax
\EndOfBibitem
\bibitem[Herbert(2015)]{Herbert:RCC:15}
J.~M. Herbert, \emph{Rev. Comp. Chem.}, 2015, \textbf{28}, 391--517\relax
\mciteBstWouldAddEndPuncttrue
\mciteSetBstMidEndSepPunct{\mcitedefaultmidpunct}
{\mcitedefaultendpunct}{\mcitedefaultseppunct}\relax
\EndOfBibitem
\bibitem[Garrett(1982)]{Garrett:JCP:82}
W.~R. Garrett, \emph{J. Chem. Phys.}, 1982, \textbf{77}, 3666--3673\relax
\mciteBstWouldAddEndPuncttrue
\mciteSetBstMidEndSepPunct{\mcitedefaultmidpunct}
{\mcitedefaultendpunct}{\mcitedefaultseppunct}\relax
\EndOfBibitem
\bibitem[Gutsev \emph{et~al.}(1997)Gutsev, Nooijen, and
  Bartlett]{Gutsev:CPL:97}
G.~L. Gutsev, M.~Nooijen and R.~J. Bartlett, \emph{Chem. Phys. Lett.}, 1997,
  \textbf{276}, 13 -- 19\relax
\mciteBstWouldAddEndPuncttrue
\mciteSetBstMidEndSepPunct{\mcitedefaultmidpunct}
{\mcitedefaultendpunct}{\mcitedefaultseppunct}\relax
\EndOfBibitem
\bibitem[Sommerfeld(2002)]{Sommerfeld:PCCP:02}
T.~Sommerfeld, \emph{Phys. Chem. Chem. Phys.}, 2002, \textbf{4},
  2511--2516\relax
\mciteBstWouldAddEndPuncttrue
\mciteSetBstMidEndSepPunct{\mcitedefaultmidpunct}
{\mcitedefaultendpunct}{\mcitedefaultseppunct}\relax
\EndOfBibitem
\bibitem[Sommerfeld \emph{et~al.}(2010)Sommerfeld, Bhattarai, Vysotskiy, and
  Cederbaum]{Sommerfeld:NaCl:10}
T.~Sommerfeld, B.~Bhattarai, V.~Vysotskiy and L.~Cederbaum, \emph{J. Chem.
  Phys.}, 2010, \textbf{133}, 114301\relax
\mciteBstWouldAddEndPuncttrue
\mciteSetBstMidEndSepPunct{\mcitedefaultmidpunct}
{\mcitedefaultendpunct}{\mcitedefaultseppunct}\relax
\EndOfBibitem
\bibitem[Voora and Jordan(2014)]{Voora:14}
V.~K. Voora and K.~D. Jordan, \emph{J. Phys. Chem. A}, 2014, \textbf{118},
  7201--7205\relax
\mciteBstWouldAddEndPuncttrue
\mciteSetBstMidEndSepPunct{\mcitedefaultmidpunct}
{\mcitedefaultendpunct}{\mcitedefaultseppunct}\relax
\EndOfBibitem
\bibitem[Voora \emph{et~al.}(2013)Voora, Cederbaum, and Jordan]{Voora:C60:2013}
V.~Voora, L.~Cederbaum and K.~Jordan, \emph{J. Phys. Chem. Lett.}, 2013,
  \textbf{4}, 849--853\relax
\mciteBstWouldAddEndPuncttrue
\mciteSetBstMidEndSepPunct{\mcitedefaultmidpunct}
{\mcitedefaultendpunct}{\mcitedefaultseppunct}\relax
\EndOfBibitem
\bibitem[Jordan and Wang(2003)]{Jordan:DBRev:03}
K.~Jordan and F.~Wang, \emph{Annu. Rev. Phys. Chem.}, 2003, \textbf{54},
  367--396\relax
\mciteBstWouldAddEndPuncttrue
\mciteSetBstMidEndSepPunct{\mcitedefaultmidpunct}
{\mcitedefaultendpunct}{\mcitedefaultseppunct}\relax
\EndOfBibitem
\bibitem[Gutowski \emph{et~al.}(1998)Gutowski, Jordan, and
  Skurski]{Gutovski:DBA:98}
M.~Gutowski, K.~Jordan and P.~Skurski, \emph{J. Phys. Chem. A}, 1998,
  \textbf{102}, 2624--2633\relax
\mciteBstWouldAddEndPuncttrue
\mciteSetBstMidEndSepPunct{\mcitedefaultmidpunct}
{\mcitedefaultendpunct}{\mcitedefaultseppunct}\relax
\EndOfBibitem
\bibitem[Fortenberry and Crawford(2011)]{Crawford:DBS:11}
R.~Fortenberry and T.~D. Crawford, \emph{J. Chem. Phys.}, 2011, \textbf{134},
  154304\relax
\mciteBstWouldAddEndPuncttrue
\mciteSetBstMidEndSepPunct{\mcitedefaultmidpunct}
{\mcitedefaultendpunct}{\mcitedefaultseppunct}\relax
\EndOfBibitem
\bibitem[Lykke \emph{et~al.}(1984)Lykke, Mead, and Lineberger]{Lykke:PRL:84}
K.~R. Lykke, R.~D. Mead and W.~C. Lineberger, \emph{Phys. Rev. Lett.}, 1984,
  \textbf{52}, 2221--2224\relax
\mciteBstWouldAddEndPuncttrue
\mciteSetBstMidEndSepPunct{\mcitedefaultmidpunct}
{\mcitedefaultendpunct}{\mcitedefaultseppunct}\relax
\EndOfBibitem
\bibitem[Jordan \emph{et~al.}(2014)Jordan, Voora, and Simons]{Simons:RevAn:14}
K.~Jordan, V.~Voora and J.~Simons, \emph{Theor. Chem. Acc.}, 2014,
  \textbf{133}, 1445\relax
\mciteBstWouldAddEndPuncttrue
\mciteSetBstMidEndSepPunct{\mcitedefaultmidpunct}
{\mcitedefaultendpunct}{\mcitedefaultseppunct}\relax
\EndOfBibitem
\bibitem[FR()]{FR}
The $(\pi\pi^*)$ excited state of the PYP chromophore is a shape resonance in
  the phenolate form and a Feshbach resonance in the carboxylate
  form\cite{Zuev:PYP:10}. An example of doubly excited anionic Feshbach
  resonance is the $(\pi)^{-1}(3s)^2$ state in ethylene.\relax
\mciteBstWouldAddEndPunctfalse
\mciteSetBstMidEndSepPunct{\mcitedefaultmidpunct}
{}{\mcitedefaultseppunct}\relax
\EndOfBibitem
\bibitem[Jordan and Burrow(1978)]{Jordan:Acetylene:78}
K.~D. Jordan and P.~D. Burrow, \emph{Acc. Chem. Res.}, 1978, \textbf{11},
  341--348\relax
\mciteBstWouldAddEndPuncttrue
\mciteSetBstMidEndSepPunct{\mcitedefaultmidpunct}
{\mcitedefaultendpunct}{\mcitedefaultseppunct}\relax
\EndOfBibitem
\bibitem[Dressler and Allan(1987)]{Dressler:Acetylene:87}
R.~Dressler and M.~Allan, \emph{J. Chem. Phys.}, 1987, \textbf{87},
  4510--4518\relax
\mciteBstWouldAddEndPuncttrue
\mciteSetBstMidEndSepPunct{\mcitedefaultmidpunct}
{\mcitedefaultendpunct}{\mcitedefaultseppunct}\relax
\EndOfBibitem
\bibitem[Schiedt and Weinkauf(1999)]{Weinkauf:benzoQ:99}
J.~Schiedt and R.~Weinkauf, \emph{J. Chem. Phys.}, 1999, \textbf{110},
  304\relax
\mciteBstWouldAddEndPuncttrue
\mciteSetBstMidEndSepPunct{\mcitedefaultmidpunct}
{\mcitedefaultendpunct}{\mcitedefaultseppunct}\relax
\EndOfBibitem
\bibitem[McCarthy \emph{et~al.}(2006)McCarthy, Gottlieb, Gupta, and
  Thaddeus]{McCarthy:AJL:06}
M.~C. McCarthy, C.~A. Gottlieb, H.~Gupta and P.~Thaddeus, \emph{Astrophys. J.
  Lett.}, 2006, \textbf{652}, L141\relax
\mciteBstWouldAddEndPuncttrue
\mciteSetBstMidEndSepPunct{\mcitedefaultmidpunct}
{\mcitedefaultendpunct}{\mcitedefaultseppunct}\relax
\EndOfBibitem
\bibitem[Br{\"u}nken \emph{et~al.}(2007)Br{\"u}nken, Gupta, Gottlieb, McCarthy,
  and Thaddeus]{Bruken:AJL:07}
S.~Br{\"u}nken, H.~Gupta, C.~A. Gottlieb, M.~C. McCarthy and P.~Thaddeus,
  \emph{Astrophys. J. Lett.}, 2007, \textbf{664}, L43\relax
\mciteBstWouldAddEndPuncttrue
\mciteSetBstMidEndSepPunct{\mcitedefaultmidpunct}
{\mcitedefaultendpunct}{\mcitedefaultseppunct}\relax
\EndOfBibitem
\bibitem[Thaddeus \emph{et~al.}(2008)Thaddeus, Gottlieb, Gupta, Br{\"u}nken,
  McCarthy, Ag{\'u}ndez, Guelin, and Cernicharo]{Thaddeus:AJ:08}
P.~Thaddeus, C.~A. Gottlieb, H.~Gupta, S.~Br{\"u}nken, M.~C. McCarthy,
  M.~Ag{\'u}ndez, M.~Guelin and J.~Cernicharo, \emph{Astrophys. J.}, 2008,
  \textbf{677}, 1132\relax
\mciteBstWouldAddEndPuncttrue
\mciteSetBstMidEndSepPunct{\mcitedefaultmidpunct}
{\mcitedefaultendpunct}{\mcitedefaultseppunct}\relax
\EndOfBibitem
\bibitem[{Cernicharo, J.} \emph{et~al.}(2007){Cernicharo, J.}, {Gu{\'e}lin,
  M.}, {Ag{\'u}ndez, M.}, {Kawaguchi, K.}, {McCarthy, M.}, and {Thaddeus,
  P.}]{Cernicharo:AA:07}
{Cernicharo, J.}, {Gu{\'e}lin, M.}, {Ag{\'u}ndez, M.}, {Kawaguchi, K.},
  {McCarthy, M.} and {Thaddeus, P.}, \emph{Astronomy \& Astrophysics}, 2007,
  \textbf{467}, L37--L40\relax
\mciteBstWouldAddEndPuncttrue
\mciteSetBstMidEndSepPunct{\mcitedefaultmidpunct}
{\mcitedefaultendpunct}{\mcitedefaultseppunct}\relax
\EndOfBibitem
\bibitem[Cernicharo \emph{et~al.}(2008)Cernicharo, Gu{\'e}lin, Agundez,
  McCarthy, and Thaddeus]{Cernicharo:AJL:08}
J.~Cernicharo, M.~Gu{\'e}lin, M.~Agundez, M.~C. McCarthy and P.~Thaddeus,
  \emph{Astrophys. J. Lett.}, 2008, \textbf{688}, L83\relax
\mciteBstWouldAddEndPuncttrue
\mciteSetBstMidEndSepPunct{\mcitedefaultmidpunct}
{\mcitedefaultendpunct}{\mcitedefaultseppunct}\relax
\EndOfBibitem
\bibitem[Ag{\'u}ndez \emph{et~al.}(2010)Ag{\'u}ndez, Cernicharo, Gu{\'e}lin,
  Kahane, Roueff, K{\l}os, Aoiz, Lique, Marcelino, Goicoechea,
  Gonz{\'a}lez-Garcia, Gottlieb, McCarthy, and Thaddeus]{Agundez:AA:10}
M.~Ag{\'u}ndez, J.~Cernicharo, M.~Gu{\'e}lin, C.~Kahane, E.~Roueff, J.~K{\l}os,
  F.~Aoiz, F.~Lique, N.~Marcelino, J.~R. Goicoechea, M.~Gonz{\'a}lez-Garcia,
  C.~Gottlieb, M.~McCarthy and P.~Thaddeus, \emph{Astronomy \& Astrophysics},
  2010, \textbf{517}, L2\relax
\mciteBstWouldAddEndPuncttrue
\mciteSetBstMidEndSepPunct{\mcitedefaultmidpunct}
{\mcitedefaultendpunct}{\mcitedefaultseppunct}\relax
\EndOfBibitem
\bibitem[Millar \emph{et~al.}(2017)Millar, Walsh, and Field]{Millar:ChemRev:17}
T.~J. Millar, C.~Walsh and T.~A. Field, \emph{Chem. Rev.}, 2017, \textbf{117},
  1765--1795\relax
\mciteBstWouldAddEndPuncttrue
\mciteSetBstMidEndSepPunct{\mcitedefaultmidpunct}
{\mcitedefaultendpunct}{\mcitedefaultseppunct}\relax
\EndOfBibitem
\bibitem[Gupta \emph{et~al.}(2007)Gupta, Br{\"u}nken, Tamassia, Gottlieb,
  McCarthy, and Thaddeus]{Gupta:AJL:07}
H.~Gupta, S.~Br{\"u}nken, F.~Tamassia, C.~A. Gottlieb, M.~C. McCarthy and
  P.~Thaddeus, \emph{Astrophys. J. Lett.}, 2007, \textbf{655}, L57\relax
\mciteBstWouldAddEndPuncttrue
\mciteSetBstMidEndSepPunct{\mcitedefaultmidpunct}
{\mcitedefaultendpunct}{\mcitedefaultseppunct}\relax
\EndOfBibitem
\bibitem[Botschwina and Oswald(2008)]{Botschwina:JCP:08}
P.~Botschwina and R.~Oswald, \emph{J. Chem. Phys.}, 2008, \textbf{129},
  044305\relax
\mciteBstWouldAddEndPuncttrue
\mciteSetBstMidEndSepPunct{\mcitedefaultmidpunct}
{\mcitedefaultendpunct}{\mcitedefaultseppunct}\relax
\EndOfBibitem
\bibitem[McCarthy and Thaddeus(2008)]{McCarthy:JCP:08}
M.~C. McCarthy and P.~Thaddeus, \emph{J. Chem. Phys.}, 2008, \textbf{129},
  054314\relax
\mciteBstWouldAddEndPuncttrue
\mciteSetBstMidEndSepPunct{\mcitedefaultmidpunct}
{\mcitedefaultendpunct}{\mcitedefaultseppunct}\relax
\EndOfBibitem
\bibitem[Gottlieb \emph{et~al.}(2007)Gottlieb, Br{\"u}nken, McCarthy, and
  Thaddeus]{Gottlieb:JCP:07}
C.~A. Gottlieb, S.~Br{\"u}nken, M.~C. McCarthy and P.~Thaddeus, \emph{J. Chem.
  Phys.}, 2007, \textbf{126}, 191101\relax
\mciteBstWouldAddEndPuncttrue
\mciteSetBstMidEndSepPunct{\mcitedefaultmidpunct}
{\mcitedefaultendpunct}{\mcitedefaultseppunct}\relax
\EndOfBibitem
\bibitem[Aoki(2000)]{Aoki:CPL:00}
K.~Aoki, \emph{Chem. Phys. Lett.}, 2000, \textbf{323}, 55 -- 58\relax
\mciteBstWouldAddEndPuncttrue
\mciteSetBstMidEndSepPunct{\mcitedefaultmidpunct}
{\mcitedefaultendpunct}{\mcitedefaultseppunct}\relax
\EndOfBibitem
\bibitem[Wang \emph{et~al.}(1995)Wang, Huang, Liu, and Zheng]{Wang:CPL:95}
C.-R. Wang, R.-B. Huang, Z.-Y. Liu and L.-S. Zheng, \emph{Chem. Phys. Lett.},
  1995, \textbf{237}, 463 -- 467\relax
\mciteBstWouldAddEndPuncttrue
\mciteSetBstMidEndSepPunct{\mcitedefaultmidpunct}
{\mcitedefaultendpunct}{\mcitedefaultseppunct}\relax
\EndOfBibitem
\bibitem[Zhan and Iwata(1996)]{Zhan:JCP:96}
C.~Zhan and S.~Iwata, \emph{J. Chem. Phys.}, 1996, \textbf{104},
  9058--9064\relax
\mciteBstWouldAddEndPuncttrue
\mciteSetBstMidEndSepPunct{\mcitedefaultmidpunct}
{\mcitedefaultendpunct}{\mcitedefaultseppunct}\relax
\EndOfBibitem
\bibitem[Coupeaud \emph{et~al.}(2008)Coupeaud, Turowski, Gronowski, Pi{\'e}tri,
  Couturier-Tamburelli, Ko{\l}os, and Aycard]{Coupeaud:JCP:08}
A.~Coupeaud, M.~Turowski, M.~Gronowski, N.~Pi{\'e}tri, I.~Couturier-Tamburelli,
  R.~Ko{\l}os and J.-P. Aycard, \emph{J. Chem. Phys.}, 2008, \textbf{128},
  154303\relax
\mciteBstWouldAddEndPuncttrue
\mciteSetBstMidEndSepPunct{\mcitedefaultmidpunct}
{\mcitedefaultendpunct}{\mcitedefaultseppunct}\relax
\EndOfBibitem
\bibitem[Stein \emph{et~al.}(2015)Stein, Weser, Schr{\"o}der, and
  Botschwina]{Stein:MP:15}
C.~Stein, O.~Weser, B.~Schr{\"o}der and P.~Botschwina, \emph{Mol. Phys.}, 2015,
  \textbf{113}, 2169--2178\relax
\mciteBstWouldAddEndPuncttrue
\mciteSetBstMidEndSepPunct{\mcitedefaultmidpunct}
{\mcitedefaultendpunct}{\mcitedefaultseppunct}\relax
\EndOfBibitem
\bibitem[Ko{\l}os \emph{et~al.}(2008)Ko{\l}os, Gronowski, and
  Botschwina]{Kolos:JCP:08}
R.~Ko{\l}os, M.~Gronowski and P.~Botschwina, \emph{J. Chem. Phys.}, 2008,
  \textbf{128}, 154305\relax
\mciteBstWouldAddEndPuncttrue
\mciteSetBstMidEndSepPunct{\mcitedefaultmidpunct}
{\mcitedefaultendpunct}{\mcitedefaultseppunct}\relax
\EndOfBibitem
\bibitem[Ito \emph{et~al.}(2014)Ito, Furukawa, Tanuma, Matsumoto, Shiromaru,
  Majima, Goto, Azuma, and Hansen]{Ito:PRL:14}
G.~Ito, T.~Furukawa, H.~Tanuma, J.~Matsumoto, H.~Shiromaru, T.~Majima, M.~Goto,
  T.~Azuma and K.~Hansen, \emph{Phys. Rev. Lett.}, 2014, \textbf{112},
  183001\relax
\mciteBstWouldAddEndPuncttrue
\mciteSetBstMidEndSepPunct{\mcitedefaultmidpunct}
{\mcitedefaultendpunct}{\mcitedefaultseppunct}\relax
\EndOfBibitem
\bibitem[Zhu \emph{et~al.}(2017)Zhu, Liu, and Wang]{Zhu:PRL:17}
G.-Z. Zhu, Y.~Liu and L.-S. Wang, \emph{Phys. Rev. Lett.}, 2017, \textbf{119},
  023002\relax
\mciteBstWouldAddEndPuncttrue
\mciteSetBstMidEndSepPunct{\mcitedefaultmidpunct}
{\mcitedefaultendpunct}{\mcitedefaultseppunct}\relax
\EndOfBibitem
\bibitem[Yzombard \emph{et~al.}(2015)Yzombard, Hamamda, Gerber, Doser, and
  Comparat]{Yzombard:PRL:15}
P.~Yzombard, M.~Hamamda, S.~Gerber, M.~Doser and D.~Comparat, \emph{Phys. Rev.
  Lett.}, 2015, \textbf{114}, 213001\relax
\mciteBstWouldAddEndPuncttrue
\mciteSetBstMidEndSepPunct{\mcitedefaultmidpunct}
{\mcitedefaultendpunct}{\mcitedefaultseppunct}\relax
\EndOfBibitem
\bibitem[Tomza(2017)]{Tomza:PCCP:17}
M.~Tomza, \emph{Phys. Chem. Chem. Phys.}, 2017, \textbf{19}, 16512--16523\relax
\mciteBstWouldAddEndPuncttrue
\mciteSetBstMidEndSepPunct{\mcitedefaultmidpunct}
{\mcitedefaultendpunct}{\mcitedefaultseppunct}\relax
\EndOfBibitem
\bibitem[Herbst and Osamura(2008)]{Herbst:AJ:08}
E.~Herbst and Y.~Osamura, \emph{Astrophys. J.}, 2008, \textbf{679}, 1670\relax
\mciteBstWouldAddEndPuncttrue
\mciteSetBstMidEndSepPunct{\mcitedefaultmidpunct}
{\mcitedefaultendpunct}{\mcitedefaultseppunct}\relax
\EndOfBibitem
\bibitem[Carelli \emph{et~al.}(2013)Carelli, Satta, Grassi, and
  Gianturco]{Carelli:AJ:13}
F.~Carelli, M.~Satta, T.~Grassi and F.~A. Gianturco, \emph{Astrophys. J.},
  2013, \textbf{774}, 97\relax
\mciteBstWouldAddEndPuncttrue
\mciteSetBstMidEndSepPunct{\mcitedefaultmidpunct}
{\mcitedefaultendpunct}{\mcitedefaultseppunct}\relax
\EndOfBibitem
\bibitem[Carelli \emph{et~al.}(2014)Carelli, Gianturco, Wester, and
  Satta]{Carelli:DBS:14}
F.~Carelli, F.~A. Gianturco, R.~Wester and M.~Satta, \emph{J. Chem. Phys.},
  2014, \textbf{141}, 054302\relax
\mciteBstWouldAddEndPuncttrue
\mciteSetBstMidEndSepPunct{\mcitedefaultmidpunct}
{\mcitedefaultendpunct}{\mcitedefaultseppunct}\relax
\EndOfBibitem
\bibitem[Satta \emph{et~al.}(2015)Satta, Gianturco, Carelli, and
  Wester]{Satta:AJ:15}
M.~Satta, F.~A. Gianturco, F.~Carelli and R.~Wester, \emph{Astrophys. J.},
  2015, \textbf{799}, 228\relax
\mciteBstWouldAddEndPuncttrue
\mciteSetBstMidEndSepPunct{\mcitedefaultmidpunct}
{\mcitedefaultendpunct}{\mcitedefaultseppunct}\relax
\EndOfBibitem
\bibitem[Khamesian \emph{et~al.}(2016)Khamesian, Douguet, {dos Santos}, Dulieu,
  Raoult, Brigg, and Kokoouline]{Douguet:CxNm:16}
M.~Khamesian, N.~Douguet, S.~{dos Santos}, O.~Dulieu, M.~Raoult, W.~Brigg and
  V.~Kokoouline, \emph{Phys. Rev. Lett.}, 2016, \textbf{117}, 123001\relax
\mciteBstWouldAddEndPuncttrue
\mciteSetBstMidEndSepPunct{\mcitedefaultmidpunct}
{\mcitedefaultendpunct}{\mcitedefaultseppunct}\relax
\EndOfBibitem
\bibitem[Douguet \emph{et~al.}(2015)Douguet, dos Santos, Raoult, Dulieu, Orel,
  and Kokoouline]{Douguet:JCP:15}
N.~Douguet, S.~F. dos Santos, M.~Raoult, O.~Dulieu, A.~E. Orel and
  V.~Kokoouline, \emph{J. Chem. Phys.}, 2015, \textbf{142}, 234309\relax
\mciteBstWouldAddEndPuncttrue
\mciteSetBstMidEndSepPunct{\mcitedefaultmidpunct}
{\mcitedefaultendpunct}{\mcitedefaultseppunct}\relax
\EndOfBibitem
\bibitem[Graupner \emph{et~al.}(2008)Graupner, Field, and
  Saunders]{Graupner:AJL:95}
K.~Graupner, T.~A. Field and G.~C. Saunders, \emph{Astrophys. J. Lett.}, 2008,
  \textbf{685}, L95\relax
\mciteBstWouldAddEndPuncttrue
\mciteSetBstMidEndSepPunct{\mcitedefaultmidpunct}
{\mcitedefaultendpunct}{\mcitedefaultseppunct}\relax
\EndOfBibitem
\bibitem[Sebastianelli and Gianturco(2012)]{Sebastianelli:EPDJ:12}
F.~Sebastianelli and F.~A. Gianturco, \emph{Eur. Phys. J. D}, 2012,
  \textbf{66}, 41\relax
\mciteBstWouldAddEndPuncttrue
\mciteSetBstMidEndSepPunct{\mcitedefaultmidpunct}
{\mcitedefaultendpunct}{\mcitedefaultseppunct}\relax
\EndOfBibitem
\bibitem[Sebastianelli and Gianturco(2010)]{Sebastianelli:EPJD:10}
F.~Sebastianelli and F.~A. Gianturco, \emph{Eur. Phys. J. D}, 2010,
  \textbf{59}, 389--398\relax
\mciteBstWouldAddEndPuncttrue
\mciteSetBstMidEndSepPunct{\mcitedefaultmidpunct}
{\mcitedefaultendpunct}{\mcitedefaultseppunct}\relax
\EndOfBibitem
\bibitem[Harrison and Tennyson(2011)]{Harrison:JPB:11}
S.~Harrison and J.~Tennyson, \emph{J. Phys. B}, 2011, \textbf{44}, 045206\relax
\mciteBstWouldAddEndPuncttrue
\mciteSetBstMidEndSepPunct{\mcitedefaultmidpunct}
{\mcitedefaultendpunct}{\mcitedefaultseppunct}\relax
\EndOfBibitem
\bibitem[Harrison and Tennyson(2012)]{Harrison:JPB:12}
S.~Harrison and J.~Tennyson, \emph{J. Phys. B}, 2012, \textbf{45}, 035204\relax
\mciteBstWouldAddEndPuncttrue
\mciteSetBstMidEndSepPunct{\mcitedefaultmidpunct}
{\mcitedefaultendpunct}{\mcitedefaultseppunct}\relax
\EndOfBibitem
\bibitem[Krylov(2008)]{Krylov:EOMRev:07}
A.~I. Krylov, \emph{Annu. Rev. Phys. Chem.}, 2008, \textbf{59}, 433--462\relax
\mciteBstWouldAddEndPuncttrue
\mciteSetBstMidEndSepPunct{\mcitedefaultmidpunct}
{\mcitedefaultendpunct}{\mcitedefaultseppunct}\relax
\EndOfBibitem
\bibitem[Sneskov and Christiansen(2012)]{Christiansen:EOMRev:11}
K.~Sneskov and O.~Christiansen, \emph{WIREs Comput. Mol. Sci.}, 2012,
  \textbf{2}, 566--584\relax
\mciteBstWouldAddEndPuncttrue
\mciteSetBstMidEndSepPunct{\mcitedefaultmidpunct}
{\mcitedefaultendpunct}{\mcitedefaultseppunct}\relax
\EndOfBibitem
\bibitem[Bartlett(2012)]{Bartlet:EOMREV:12}
R.~Bartlett, \emph{WIREs Comput. Mol. Sci.}, 2012, \textbf{2}, 126--138\relax
\mciteBstWouldAddEndPuncttrue
\mciteSetBstMidEndSepPunct{\mcitedefaultmidpunct}
{\mcitedefaultendpunct}{\mcitedefaultseppunct}\relax
\EndOfBibitem
\bibitem[Fortenberry(2015)]{Fortenberry:JPCA:15}
R.~C. Fortenberry, \emph{J. Phys. Chem. A}, 2015, \textbf{119},
  9941--9953\relax
\mciteBstWouldAddEndPuncttrue
\mciteSetBstMidEndSepPunct{\mcitedefaultmidpunct}
{\mcitedefaultendpunct}{\mcitedefaultseppunct}\relax
\EndOfBibitem
\bibitem[Kumar \emph{et~al.}(2013)Kumar, Hauser, Jindra, Best, Roucka, Geppert,
  Millar, and Wester]{Wester:interstellarAnDestruct:13}
S.~S. Kumar, D.~Hauser, R.~Jindra, T.~Best, S.~Roucka, W.~D. Geppert, T.~J.
  Millar and R.~Wester, \emph{Astrophys. J.}, 2013, \textbf{776}, 25\relax
\mciteBstWouldAddEndPuncttrue
\mciteSetBstMidEndSepPunct{\mcitedefaultmidpunct}
{\mcitedefaultendpunct}{\mcitedefaultseppunct}\relax
\EndOfBibitem
\bibitem[Bradforth \emph{et~al.}(1993)Bradforth, Kim, Arnold, and
  Neumark]{Bradforth:JCP:98}
S.~E. Bradforth, E.~H. Kim, D.~W. Arnold and D.~M. Neumark, \emph{J. Chem.
  Phys.}, 1993, \textbf{98}, 800--810\relax
\mciteBstWouldAddEndPuncttrue
\mciteSetBstMidEndSepPunct{\mcitedefaultmidpunct}
{\mcitedefaultendpunct}{\mcitedefaultseppunct}\relax
\EndOfBibitem
\bibitem[Yen \emph{et~al.}(2010)Yen, Garand, Shreve, and Neumark]{Yen:JCPA:10}
T.~A. Yen, E.~Garand, A.~T. Shreve and D.~M. Neumark, \emph{J. Phys. Chem. A},
  2010, \textbf{114}, 3215--3220\relax
\mciteBstWouldAddEndPuncttrue
\mciteSetBstMidEndSepPunct{\mcitedefaultmidpunct}
{\mcitedefaultendpunct}{\mcitedefaultseppunct}\relax
\EndOfBibitem
\bibitem[Grutter \emph{et~al.}(1999)Grutter, Wyss, and Maier]{Grutter:JCP:99}
M.~Grutter, M.~Wyss and J.~P. Maier, \emph{J. Chem. Phys.}, 1999, \textbf{110},
  1492--1496\relax
\mciteBstWouldAddEndPuncttrue
\mciteSetBstMidEndSepPunct{\mcitedefaultmidpunct}
{\mcitedefaultendpunct}{\mcitedefaultseppunct}\relax
\EndOfBibitem
\bibitem[Jagau \emph{et~al.}(2017)Jagau, Bravaya, and Krylov]{KrylovResReview}
T.-C. Jagau, K.~B. Bravaya and A.~I. Krylov, \emph{Annu. Rev. Phys. Chem.},
  2017, \textbf{68}, 525--553\relax
\mciteBstWouldAddEndPuncttrue
\mciteSetBstMidEndSepPunct{\mcitedefaultmidpunct}
{\mcitedefaultendpunct}{\mcitedefaultseppunct}\relax
\EndOfBibitem
\bibitem[Reisler and Krylov(2009)]{RydbergReview09}
H.~Reisler and A.~I. Krylov, \emph{Int. Rev. Phys. Chem.}, 2009, \textbf{28},
  267--308\relax
\mciteBstWouldAddEndPuncttrue
\mciteSetBstMidEndSepPunct{\mcitedefaultmidpunct}
{\mcitedefaultendpunct}{\mcitedefaultseppunct}\relax
\EndOfBibitem
\bibitem[Krylov(2017)]{Krylov:OSRev}
A.~I. Krylov, \emph{Reviews in Comp. Chem.}, J. Wiley \& Sons, 2017, vol.~30,
  ch.~4, pp. 151--224\relax
\mciteBstWouldAddEndPuncttrue
\mciteSetBstMidEndSepPunct{\mcitedefaultmidpunct}
{\mcitedefaultendpunct}{\mcitedefaultseppunct}\relax
\EndOfBibitem
\bibitem[Zuev \emph{et~al.}(2014)Zuev, Jagau, Bravaya, Epifanovsky, Shao,
  Sundstrom, Head-Gordon, and Krylov]{Zuev:CAP:14}
D.~Zuev, T.-C. Jagau, K.~B. Bravaya, E.~Epifanovsky, Y.~Shao, E.~Sundstrom,
  M.~Head-Gordon and A.~I. Krylov, \emph{J. Chem. Phys.}, 2014, \textbf{141},
  024102\relax
\mciteBstWouldAddEndPuncttrue
\mciteSetBstMidEndSepPunct{\mcitedefaultmidpunct}
{\mcitedefaultendpunct}{\mcitedefaultseppunct}\relax
\EndOfBibitem
\bibitem[Jagau \emph{et~al.}(2014)Jagau, Zuev, Bravaya, Epifanovsky, and
  Krylov]{Jagau:CAP:13}
T.-C. Jagau, D.~Zuev, K.~B. Bravaya, E.~Epifanovsky and A.~I. Krylov, \emph{J.
  Phys. Chem. Lett.}, 2014, \textbf{5}, 310--315\relax
\mciteBstWouldAddEndPuncttrue
\mciteSetBstMidEndSepPunct{\mcitedefaultmidpunct}
{\mcitedefaultendpunct}{\mcitedefaultseppunct}\relax
\EndOfBibitem
\bibitem[Jagau and Krylov(2014)]{Jagau:PES:14}
T.-C. Jagau and A.~I. Krylov, \emph{J. Phys. Chem. Lett.}, 2014, \textbf{5},
  3078--3085\relax
\mciteBstWouldAddEndPuncttrue
\mciteSetBstMidEndSepPunct{\mcitedefaultmidpunct}
{\mcitedefaultendpunct}{\mcitedefaultseppunct}\relax
\EndOfBibitem
\bibitem[Ghosh \emph{et~al.}(2012)Ghosh, Vaval, and Pal]{Pal:CAPEOMCC:12}
A.~Ghosh, N.~Vaval and S.~Pal, \emph{J. Chem. Phys.}, 2012, \textbf{136},
  234110\relax
\mciteBstWouldAddEndPuncttrue
\mciteSetBstMidEndSepPunct{\mcitedefaultmidpunct}
{\mcitedefaultendpunct}{\mcitedefaultseppunct}\relax
\EndOfBibitem
\bibitem[Ehara and Sommerfeld(2012)]{Sommerfeld:CAP-SACCI:12}
M.~Ehara and T.~Sommerfeld, \emph{Chem. Phys. Lett.}, 2012, \textbf{537},
  107--112\relax
\mciteBstWouldAddEndPuncttrue
\mciteSetBstMidEndSepPunct{\mcitedefaultmidpunct}
{\mcitedefaultendpunct}{\mcitedefaultseppunct}\relax
\EndOfBibitem
\bibitem[Jagau \emph{et~al.}(2015)Jagau, Dao, Holtgrewe, Krylov, and
  Mabbs]{Jagau:AgCuF:15}
T.-C. Jagau, D.~B. Dao, N.~S. Holtgrewe, A.~I. Krylov and R.~Mabbs, \emph{J.
  Phys. Chem. Lett.}, 2015, \textbf{6}, 2786--2793\relax
\mciteBstWouldAddEndPuncttrue
\mciteSetBstMidEndSepPunct{\mcitedefaultmidpunct}
{\mcitedefaultendpunct}{\mcitedefaultseppunct}\relax
\EndOfBibitem
\bibitem[Bartlett and Musial(2007)]{Bartlett:CC_review:07}
R.~Bartlett and M.~Musial, \emph{Rev. Mod. Phys.}, 2007, \textbf{79},
  291--352\relax
\mciteBstWouldAddEndPuncttrue
\mciteSetBstMidEndSepPunct{\mcitedefaultmidpunct}
{\mcitedefaultendpunct}{\mcitedefaultseppunct}\relax
\EndOfBibitem
\bibitem[Riss and Meyer(1993)]{Meyer:CAP:93}
U.~V. Riss and H.-D. Meyer, \emph{J. Phys. B}, 1993, \textbf{26},
  4503--4536\relax
\mciteBstWouldAddEndPuncttrue
\mciteSetBstMidEndSepPunct{\mcitedefaultmidpunct}
{\mcitedefaultendpunct}{\mcitedefaultseppunct}\relax
\EndOfBibitem
\bibitem[Santra and Cederbaum(2002)]{Santra:CAPREV:02}
R.~Santra and L.~Cederbaum, \emph{Phys. Rep.}, 2002, \textbf{368}, 1--117\relax
\mciteBstWouldAddEndPuncttrue
\mciteSetBstMidEndSepPunct{\mcitedefaultmidpunct}
{\mcitedefaultendpunct}{\mcitedefaultseppunct}\relax
\EndOfBibitem
\bibitem[Muga \emph{et~al.}(2004)Muga, Palao, Navarro, and
  Egusquiza]{Muga:CAP:04}
J.~Muga, J.~Palao, B.~Navarro and I.~Egusquiza, \emph{Phys. Rep.}, 2004,
  \textbf{395}, 357--426\relax
\mciteBstWouldAddEndPuncttrue
\mciteSetBstMidEndSepPunct{\mcitedefaultmidpunct}
{\mcitedefaultendpunct}{\mcitedefaultseppunct}\relax
\EndOfBibitem
\bibitem[Sommerfeld and Meyer(2002)]{Sommerfeld:CSRev:02}
T.~Sommerfeld and H.-D. Meyer, \emph{J. Phys. B}, 2002, \textbf{35},
  1841--1863\relax
\mciteBstWouldAddEndPuncttrue
\mciteSetBstMidEndSepPunct{\mcitedefaultmidpunct}
{\mcitedefaultendpunct}{\mcitedefaultseppunct}\relax
\EndOfBibitem
\bibitem[Santra and Cederbaum(2002)]{Cederbaum:csADC:02}
R.~Santra and L.~Cederbaum, \emph{J. Chem. Phys.}, 2002, \textbf{117},
  5511--5521\relax
\mciteBstWouldAddEndPuncttrue
\mciteSetBstMidEndSepPunct{\mcitedefaultmidpunct}
{\mcitedefaultendpunct}{\mcitedefaultseppunct}\relax
\EndOfBibitem
\bibitem[Sommerfeld and Santra(2001)]{Sommerfeld:CAPCI:01}
T.~Sommerfeld and R.~Santra, \emph{Int. J. Quant. Chem.}, 2001, \textbf{82},
  218--226\relax
\mciteBstWouldAddEndPuncttrue
\mciteSetBstMidEndSepPunct{\mcitedefaultmidpunct}
{\mcitedefaultendpunct}{\mcitedefaultseppunct}\relax
\EndOfBibitem
\bibitem[Zhou and Ernzerhof(2012)]{Ernzerhof:CAPDFT:12}
Y.~Zhou and M.~Ernzerhof, \emph{J. Phys. Chem. Lett.}, 2012, \textbf{3},
  1916--1920\relax
\mciteBstWouldAddEndPuncttrue
\mciteSetBstMidEndSepPunct{\mcitedefaultmidpunct}
{\mcitedefaultendpunct}{\mcitedefaultseppunct}\relax
\EndOfBibitem
\bibitem[Riss and Meyer(1998)]{Meyer:CAP:98}
U.~Riss and H.-D. Meyer, \emph{J. Phys. B}, 1998, \textbf{31}, 2279--2304\relax
\mciteBstWouldAddEndPuncttrue
\mciteSetBstMidEndSepPunct{\mcitedefaultmidpunct}
{\mcitedefaultendpunct}{\mcitedefaultseppunct}\relax
\EndOfBibitem
\bibitem[CAP()]{CAPonset}
Note that in EOM-EA calculations, the reference corresponds to the neutral
  state, whereas in EOM-EE it is the ground anionic state.\relax
\mciteBstWouldAddEndPunctfalse
\mciteSetBstMidEndSepPunct{\mcitedefaultmidpunct}
{}{\mcitedefaultseppunct}\relax
\EndOfBibitem
\bibitem[Jagau and Krylov(2016)]{Jagau:Dyson:16}
T.-C. Jagau and A.~I. Krylov, \emph{J. Chem. Phys.}, 2016, \textbf{144},
  054113\relax
\mciteBstWouldAddEndPuncttrue
\mciteSetBstMidEndSepPunct{\mcitedefaultmidpunct}
{\mcitedefaultendpunct}{\mcitedefaultseppunct}\relax
\EndOfBibitem
\bibitem[Gozem \emph{et~al.}(2015)Gozem, Gunina, Ichino, Osborn, Stanton, and
  Krylov]{Gozem:Dyson:15}
S.~Gozem, A.~O. Gunina, T.~Ichino, D.~L. Osborn, J.~F. Stanton and A.~I.
  Krylov, \emph{J. Phys. Chem. Lett.}, 2015, \textbf{6}, 4532--4540\relax
\mciteBstWouldAddEndPuncttrue
\mciteSetBstMidEndSepPunct{\mcitedefaultmidpunct}
{\mcitedefaultendpunct}{\mcitedefaultseppunct}\relax
\EndOfBibitem
\bibitem[Kendall \emph{et~al.}(1992)Kendall, T.H.~Dunning, and
  Harrison]{Dunning:92:augccpvxz}
R.~Kendall, J.~T.H.~Dunning and R.~Harrison, \emph{J. Chem. Phys.}, 1992,
  \textbf{96}, 6796--6806\relax
\mciteBstWouldAddEndPuncttrue
\mciteSetBstMidEndSepPunct{\mcitedefaultmidpunct}
{\mcitedefaultendpunct}{\mcitedefaultseppunct}\relax
\EndOfBibitem
\bibitem[{Shao, Y.; Gan, Z.; Epifanovsky, E.; Gilbert, A.T.B.; Wormit, M.;
  Kussmann, J.; Lange, A.W.; Behn, A.; Deng, J.; Feng, X., et
  al.}(2015)]{qchem_2014s}
{Shao, Y.; Gan, Z.; Epifanovsky, E.; Gilbert, A.T.B.; Wormit, M.; Kussmann, J.;
  Lange, A.W.; Behn, A.; Deng, J.; Feng, X., et al.}, \emph{Mol. Phys.}, 2015,
  \textbf{113}, 184--215\relax
\mciteBstWouldAddEndPuncttrue
\mciteSetBstMidEndSepPunct{\mcitedefaultmidpunct}
{\mcitedefaultendpunct}{\mcitedefaultseppunct}\relax
\EndOfBibitem
\bibitem[Krylov and Gill(2013)]{qchem_feature}
A.~I. Krylov and P.~M.~W. Gill, \emph{WIREs Comput. Mol. Sci.}, 2013,
  \textbf{3}, 317--326\relax
\mciteBstWouldAddEndPuncttrue
\mciteSetBstMidEndSepPunct{\mcitedefaultmidpunct}
{\mcitedefaultendpunct}{\mcitedefaultseppunct}\relax
\EndOfBibitem
\bibitem[Graupner \emph{et~al.}(2006)Graupner, Merrigan, Field, Youngs, and
  Marr]{Graupner:NJP:06}
K.~Graupner, T.~L. Merrigan, T.~A. Field, T.~G.~A. Youngs and P.~C. Marr,
  \emph{New. J. Phys.}, 2006, \textbf{8}, 117\relax
\mciteBstWouldAddEndPuncttrue
\mciteSetBstMidEndSepPunct{\mcitedefaultmidpunct}
{\mcitedefaultendpunct}{\mcitedefaultseppunct}\relax
\EndOfBibitem
\bibitem[Leach(2012)]{Leach:MNRAS:12}
S.~Leach, \emph{Mon. N. Roy. Astr. Soc.}, 2012, \textbf{421}, 1325--1330\relax
\mciteBstWouldAddEndPuncttrue
\mciteSetBstMidEndSepPunct{\mcitedefaultmidpunct}
{\mcitedefaultendpunct}{\mcitedefaultseppunct}\relax
\EndOfBibitem
\bibitem[Casavecchia \emph{et~al.}(2002)Casavecchia, Balucani, Cartechini,
  Capozza, Bergeat, and Volpi]{Casavecchia:FD:02}
P.~Casavecchia, N.~Balucani, L.~Cartechini, G.~Capozza, A.~Bergeat and G.~G.
  Volpi, \emph{Faraday Discuss.}, 2002, \textbf{119}, 27--49\relax
\mciteBstWouldAddEndPuncttrue
\mciteSetBstMidEndSepPunct{\mcitedefaultmidpunct}
{\mcitedefaultendpunct}{\mcitedefaultseppunct}\relax
\EndOfBibitem
\bibitem[White \emph{et~al.}(2015)White, Head-Gordon, and
  McCurdy]{Head-Gordon:CS:15}
A.~White, M.~Head-Gordon and C.~McCurdy, \emph{J. Chem. Phys.}, 2015,
  \textbf{142}, 054103\relax
\mciteBstWouldAddEndPuncttrue
\mciteSetBstMidEndSepPunct{\mcitedefaultmidpunct}
{\mcitedefaultendpunct}{\mcitedefaultseppunct}\relax
\EndOfBibitem
\bibitem[White \emph{et~al.}(2017)White, Epifanovsky, McCurdy, and
  Head-Gordon]{White:17b}
A.~F. White, E.~Epifanovsky, C.~W. McCurdy and M.~Head-Gordon, \emph{J. Chem.
  Phys.}, 2017, \textbf{146}, 234107\relax
\mciteBstWouldAddEndPuncttrue
\mciteSetBstMidEndSepPunct{\mcitedefaultmidpunct}
{\mcitedefaultendpunct}{\mcitedefaultseppunct}\relax
\EndOfBibitem
\bibitem[Ha and Zumofen(1980)]{Ha:MP:80}
T.-K. Ha and G.~Zumofen, \emph{Mol. Phys.}, 1980, \textbf{40}, 445--454\relax
\mciteBstWouldAddEndPuncttrue
\mciteSetBstMidEndSepPunct{\mcitedefaultmidpunct}
{\mcitedefaultendpunct}{\mcitedefaultseppunct}\relax
\EndOfBibitem
\bibitem[Pol{\'a}k and Fi{\'s}er(2002)]{Polak:JMST:02}
R.~Pol{\'a}k and J.~Fi{\'s}er, \emph{J. Molec. Struct. (THEOCHEM)}, 2002,
  \textbf{584}, 69 -- 77\relax
\mciteBstWouldAddEndPuncttrue
\mciteSetBstMidEndSepPunct{\mcitedefaultmidpunct}
{\mcitedefaultendpunct}{\mcitedefaultseppunct}\relax
\EndOfBibitem
\bibitem[Musial(2005)]{Musial:MP:05}
M.~Musial, \emph{Mol. Phys.}, 2005, \textbf{103}, 2055--2060\relax
\mciteBstWouldAddEndPuncttrue
\mciteSetBstMidEndSepPunct{\mcitedefaultmidpunct}
{\mcitedefaultendpunct}{\mcitedefaultseppunct}\relax
\EndOfBibitem
\bibitem[Partridge \emph{et~al.}(1990)Partridge, Langhoff, and
  Jr.]{Partridge:JCP:90}
H.~Partridge, S.~R. Langhoff and C.~W.~B. Jr., \emph{J. Chem. Phys.}, 1990,
  \textbf{93}, 7179--7186\relax
\mciteBstWouldAddEndPuncttrue
\mciteSetBstMidEndSepPunct{\mcitedefaultmidpunct}
{\mcitedefaultendpunct}{\mcitedefaultseppunct}\relax
\EndOfBibitem
\bibitem[Ermler \emph{et~al.}(1982)Ermler, McLean, and Mulliken]{Ermler:JPC:82}
W.~C. Ermler, A.~D. McLean and R.~S. Mulliken, \emph{J. Phys. Chem.}, 1982,
  \textbf{86}, 1305--1314\relax
\mciteBstWouldAddEndPuncttrue
\mciteSetBstMidEndSepPunct{\mcitedefaultmidpunct}
{\mcitedefaultendpunct}{\mcitedefaultseppunct}\relax
\EndOfBibitem
\bibitem[R.~Zahradn{\'i}k and Panc{\'i}{\v r}(1971)]{Zahradnik:CCCC:71}
I.~T. R.~Zahradn{\'i}k and J.~Panc{\'i}{\v r}, \emph{Collect. Czech. Chem.
  Commun.}, 1971, \textbf{36}, 2867--2880\relax
\mciteBstWouldAddEndPuncttrue
\mciteSetBstMidEndSepPunct{\mcitedefaultmidpunct}
{\mcitedefaultendpunct}{\mcitedefaultseppunct}\relax
\EndOfBibitem
\bibitem[Bardsley \emph{et~al.}(1967)Bardsley, Mandl, and Wood]{Bardsley:67}
J.~Bardsley, F.~Mandl and A.~Wood, \emph{Chem. Phys. Lett.}, 1967, \textbf{1},
  359 -- 362\relax
\mciteBstWouldAddEndPuncttrue
\mciteSetBstMidEndSepPunct{\mcitedefaultmidpunct}
{\mcitedefaultendpunct}{\mcitedefaultseppunct}\relax
\EndOfBibitem
\bibitem[Krauss and Mies(1970)]{Krauss:70}
M.~Krauss and F.~H. Mies, \emph{Phys. Rev. A}, 1970, \textbf{1},
  1592--1598\relax
\mciteBstWouldAddEndPuncttrue
\mciteSetBstMidEndSepPunct{\mcitedefaultmidpunct}
{\mcitedefaultendpunct}{\mcitedefaultseppunct}\relax
\EndOfBibitem
\bibitem[DOn()]{DOnorms}
Complex values of the norms of Dyson orbitals\cite{Jagau:Dyson:16} are the
  consequence of c-norm used in non-Hermitian extensions of quantum
  mechanics.\relax
\mciteBstWouldAddEndPunctfalse
\mciteSetBstMidEndSepPunct{\mcitedefaultmidpunct}
{}{\mcitedefaultseppunct}\relax
\EndOfBibitem
\bibitem[Crawford(1971)]{Crawford:DBS:70}
O.~Crawford, \emph{Mol. Phys.}, 1971, \textbf{20}, 585--591\relax
\mciteBstWouldAddEndPuncttrue
\mciteSetBstMidEndSepPunct{\mcitedefaultmidpunct}
{\mcitedefaultendpunct}{\mcitedefaultseppunct}\relax
\EndOfBibitem
\bibitem[DBS()]{DBS}
We note that in order for the Davidson procedure to find these states, which
  have highly non-Koopmans character, it was necessary to use tighter
  convergence thresholds ($10^{-6}$) and a much larger than usual guess space
  (we used 120 guess vectors).\relax
\mciteBstWouldAddEndPunctfalse
\mciteSetBstMidEndSepPunct{\mcitedefaultmidpunct}
{}{\mcitedefaultseppunct}\relax
\EndOfBibitem
\bibitem[Jelisavcic \emph{et~al.}(2003)Jelisavcic, Panajotovic, and
  Buckman]{Jelisavcic:PRL:03}
M.~Jelisavcic, R.~Panajotovic and S.~J. Buckman, \emph{Phys. Rev. Lett.}, 2003,
  \textbf{90}, 203201\relax
\mciteBstWouldAddEndPuncttrue
\mciteSetBstMidEndSepPunct{\mcitedefaultmidpunct}
{\mcitedefaultendpunct}{\mcitedefaultseppunct}\relax
\EndOfBibitem
\bibitem[Pedersen \emph{et~al.}(1998)Pedersen, Djuri\ifmmode~\acute{c}\else
  \'{c}\fi{}, Jensen, Kella, Safvan, Vejby-Christensen, and
  Andersen]{Pedersen:PRL:98}
H.~B. Pedersen, N.~Djuri\ifmmode~\acute{c}\else \'{c}\fi{}, M.~J. Jensen,
  D.~Kella, C.~P. Safvan, L.~Vejby-Christensen and L.~H. Andersen, \emph{Phys.
  Rev. Lett.}, 1998, \textbf{81}, 5302--5305\relax
\mciteBstWouldAddEndPuncttrue
\mciteSetBstMidEndSepPunct{\mcitedefaultmidpunct}
{\mcitedefaultendpunct}{\mcitedefaultseppunct}\relax
\EndOfBibitem
\bibitem[Haughey \emph{et~al.}(2012)Haughey, Field, Langer, Shuman, Miller,
  Friedman, and Viggiano]{Haughey:JCP:12}
S.~A. Haughey, T.~A. Field, J.~Langer, N.~S. Shuman, T.~M. Miller, J.~F.
  Friedman and A.~A. Viggiano, \emph{J. Chem. Phys.}, 2012, \textbf{137},
  054310\relax
\mciteBstWouldAddEndPuncttrue
\mciteSetBstMidEndSepPunct{\mcitedefaultmidpunct}
{\mcitedefaultendpunct}{\mcitedefaultseppunct}\relax
\EndOfBibitem
\bibitem[Allan(1982)]{Allan:HCA:82}
M.~Allan, \emph{Helv. Chim. Acta}, 1982, \textbf{65}, 2008--2023\relax
\mciteBstWouldAddEndPuncttrue
\mciteSetBstMidEndSepPunct{\mcitedefaultmidpunct}
{\mcitedefaultendpunct}{\mcitedefaultseppunct}\relax
\EndOfBibitem
\bibitem[Turowski \emph{et~al.}(2008)Turowski, Gronowski, Boy\'{e}-P\'{e}ronne,
  Douin, Mon\'{e}ron, Cr\'{e}pin, and Ko{\l}os]{Turowski:08}
M.~Turowski, M.~Gronowski, S.~Boy\'{e}-P\'{e}ronne, S.~Douin, L.~Mon\'{e}ron,
  C.~Cr\'{e}pin and R.~Ko{\l}os, \emph{J. Chem. Phys.}, 2008, \textbf{128},
  164304\relax
\mciteBstWouldAddEndPuncttrue
\mciteSetBstMidEndSepPunct{\mcitedefaultmidpunct}
{\mcitedefaultendpunct}{\mcitedefaultseppunct}\relax
\EndOfBibitem
\bibitem[Turowski \emph{et~al.}(2010)Turowski, Cr\'{e}pin, Gronowski,
  Guillemin, Coupeaud, Couturier-Tamburelli, Pi\'{e}tri, and
  Ko{\l}os]{Turowski:10}
M.~Turowski, C.~Cr\'{e}pin, M.~Gronowski, J.-C. Guillemin, A.~Coupeaud,
  I.~Couturier-Tamburelli, N.~Pi\'{e}tri and R.~Ko{\l}os, \emph{J. Chem.
  Phys.}, 2010, \textbf{133}, 074310\relax
\mciteBstWouldAddEndPuncttrue
\mciteSetBstMidEndSepPunct{\mcitedefaultmidpunct}
{\mcitedefaultendpunct}{\mcitedefaultseppunct}\relax
\EndOfBibitem
\bibitem[Couturier-Tamburelli \emph{et~al.}(2014)Couturier-Tamburelli,
  Pi\'{e}tri, Cr\'{e}pin, Turowski, Guillemin, and Ko{\l}os]{Couturier:14}
I.~Couturier-Tamburelli, N.~Pi\'{e}tri, C.~Cr\'{e}pin, M.~Turowski, J.-C.
  Guillemin and R.~Ko{\l}os, \emph{J. Chem. Phys.}, 2014, \textbf{140},
  044329\relax
\mciteBstWouldAddEndPuncttrue
\mciteSetBstMidEndSepPunct{\mcitedefaultmidpunct}
{\mcitedefaultendpunct}{\mcitedefaultseppunct}\relax
\EndOfBibitem
\bibitem[Zhao \emph{et~al.}(1996)Zhao, de~Beer, Xu, Taylor, and
  Neumark]{Zhao:96}
Y.~Zhao, E.~de~Beer, C.~Xu, T.~Taylor and D.~M. Neumark, \emph{J. Chem. Phys.},
  1996, \textbf{105}, 4905--4919\relax
\mciteBstWouldAddEndPuncttrue
\mciteSetBstMidEndSepPunct{\mcitedefaultmidpunct}
{\mcitedefaultendpunct}{\mcitedefaultseppunct}\relax
\EndOfBibitem
\bibitem[Tulej \emph{et~al.}(2000)Tulej, Fulara, Sobolewski, Jungen, and
  Maier]{Tulej:00}
M.~Tulej, J.~Fulara, A.~Sobolewski, M.~Jungen and J.~P. Maier, \emph{J. Chem.
  Phys.}, 2000, \textbf{112}, 3747--3753\relax
\mciteBstWouldAddEndPuncttrue
\mciteSetBstMidEndSepPunct{\mcitedefaultmidpunct}
{\mcitedefaultendpunct}{\mcitedefaultseppunct}\relax
\EndOfBibitem
\bibitem[Zuev \emph{et~al.}(2011)Zuev, Bravaya, Crawford, Lindh, and
  Krylov]{Zuev:PYP:10}
D.~Zuev, K.~B. Bravaya, T.~D. Crawford, R.~Lindh and A.~I. Krylov, \emph{J.
  Chem. Phys.}, 2011, \textbf{134}, 034310\relax
\mciteBstWouldAddEndPuncttrue
\mciteSetBstMidEndSepPunct{\mcitedefaultmidpunct}
{\mcitedefaultendpunct}{\mcitedefaultseppunct}\relax
\EndOfBibitem
\end{mcitethebibliography}
\bibliographystyle{rsc}

\end{document}